\documentclass[twocolumn]{aastex63}
\usepackage{amsmath,  amssymb, url, graphicx}
\renewcommand{\vec}[1]{\pmb{#1}}

\newcommand{\beq}{\begin{equation}}
\newcommand{\eeq}{\end{equation}}
\newcommand{\tvs}{\tilde{v}_s}
\newcommand{\T}{{\cal T}}
\newcommand{\tauc}{\tau}
\newcommand{\lQ}{\ell_0}
\newcommand{\AQ}{A_0}
\newcommand{\lperp}{\ell_{\perp}}
\newcommand{\FA}{F}
\newcommand{\M}{{\cal M}}
\newcommand{\ff}{\zeta}

\accepted{May 14, 2020}
\submitjournal{ApJ}

\shorttitle{A Quake Quenching the Vela Pulsar}

\begin{document}

\title{A Quake Quenching the Vela Pulsar}

\correspondingauthor{Ashley Bransgrove}
\email{ashley.bransgrove@columbia.edu}

\author[0000-0002-9711-9424]{Ashley Bransgrove}
\affil{Physics Department and Columbia Astrophysics Laboratory, Columbia University, 538 West 120th Street, New York, NY 10027}

\author{Andrei M. Beloborodov}
\affil{Physics Department and Columbia Astrophysics Laboratory, Columbia University, 538 West 120th Street, New York, NY 10027}
\affil{Max Planck Institute for Astrophysics, Karl-Schwarzschild-Str. 1, D-85741, Garching, Germany}

\author{Yuri Levin}
\affil{Physics Department and Columbia Astrophysics Laboratory, Columbia University, 538 West 120th Street, New York, NY 10027}
\affil{Center for Computational Astrophysics, Flatiron Institute, 162 5th Avenue, 6th floor, New York, NY 10010
}



\begin{abstract}
The remarkable null pulse coincident with the 2016 glitch in Vela rotation indicates a dynamical event involving the crust and the magnetosphere of the neutron star. We propose that a crustal quake associated with the glitch strongly disturbed the Vela magnetosphere and thus interrupted its radio emission. We present the first global numerical simulations of a neutron starquake. Our code resolves the elasto-dynamics of the entire crust and follows the evolution of Alfv\'en waves excited in the magnetosphere. We observe Rayleigh surface waves propagating away from the epicenter of the quake, around the circumference of the crust --- an instance of the so-called whispering gallery modes. The Rayleigh waves set the initial spatial scale of the magnetospheric disturbance. Once launched, the Alfv\'en waves bounce in the closed magnetosphere, become de-phased, and generate strong electric currents, capable of igniting electric discharge. Most likely, the discharge floods the magnetosphere with electron-positron plasma, quenching the radio emission. We find that the observed $\sim 0.2$ s disturbance is consistent with the damping time of the crustal waves if the crust is magnetically coupled to the superconducting core of the neutron star. The quake is expected to produce a weak X-ray burst of short duration.
\end{abstract}


\section{Introduction}
\subsection{Glitches}
Pulsars are highly stable rotators, which slowly spin down. However, 
they show two types of irregularity dubbed timing noise and glitches. Timing noise is the slow stochastic deviation from regular spin-down, most prominent in young pulsars [\cite{hobbs_analysis_2010}, \cite{lyne_switched_2010}].
A glitch is a sudden increase in the spin frequency $\nu$, sometimes accompanied by a change in the spin-down rate $\dot{\nu}$.

The first pulsar glitch was observed in the Vela pulsar \citep{radhakrishnan_detection_1969},
and by now
there are more than 520 recorded glitches in 180 pulsars \citep{manchester_pulsar_2018}
with glitch magnitude (relative frequency change) ranging from $\Delta\nu / \nu \approx 10^{-12}$ to $\Delta\nu / \nu \approx 10^{-5}$ \citep{espinoza_study_2011}. The so-called `Crab-like' pulsars feature strong jumps in spin-down with $\Delta\dot{\nu}/\dot{\nu}\gg\Delta\nu/\nu$, power-law glitch-size distributions, and exponential wait-time distributions \citep{melatos_avalanche_2008}. 
The so-called `Vela-like' pulsars glitch quasi-periodically, with consistently large magnitude \citep{espinoza_study_2011}.

The standard theoretical picture of a pulsar glitch involves a sudden transfer of angular momentum to the crust due to the catastrophic unpinning of superfluid vorticity \citep{anderson_pulsar_1975}. In this picture,
the crust
(ion lattice)
spins down due to external torques
while the rotation of the crustal neutron superfluid remains unchanged as long as its vorticity (quantized vortices) is pinned to the lattice. When the rotation mismatch
builds up to some 
threshold, 
many vortices are unpinned simultaneously and migrate away from the axis of rotation,
spinning down the superfluid and spinning up the crust, 
thus bringing the two components closer to corotation.

Quakes have been proposed in the past as a possible mechanism for triggering the glitch [\cite{ruderman_crust-breaking_1976},  \cite{alpar_postglitch_1994}, \cite{link_thermally_1996}, \cite{larson_simulations_2002}, \cite{eichler_dynamical_2010}]. Quakes are expected to occur when the crust is stressed beyond a critical strain $\sim 0.1$, leading to its mechanical failure \citep{horowitz_breaking_2009}. However, there exists no compelling reason why such large stresses should ever build up in the crusts of typical pulsars, which are relatively weakly magnetized and slowly spinning. Therefore, other ideas for the glitch trigger were explored (see \cite{haskell_models_2015} for a review). Nevertheless, in this paper we argue that the 2016 glitch in the Vela pulsar, and the accompanying major magnetospheric transient observed by \cite{palfreyman_alteration_2018}, were triggered by a quake. 

\subsection{The 2016 December Vela Glitch}
On 2016 December 12 a glitch of magnitude $\Delta\nu / \nu =1.431\times 10^{-6}$ was observed in the Vela pulsar (PSR J0835-4510) 
with the 26 m telescope at Mount Pleasant, Tasmania, and the 30 m telescope at Ceduna, South Australia \citep{palfreyman_alteration_2018}. For the first time, each single radio pulse was recorded during the glitch, and the pulse shape was seen to change dramatically. First, a broad pulse was detected, followed by a single null (missing) pulse. The following two pulses showed an unusually low linear polarization. \cite{ashton_rotational_2019} constrained the rise time of the glitch to be less than 12.6 s. Additionally they found evidence for a slow-down of the pulsar immediately before the spin-up glitch. 

Detection of the  radiative feature accompanying the 2016 Vela glitch was challenging because of its very short duration (two pulses, $\sim$0.2 s) and no subsequent long-term change in the pulse shape. This is different from the known behavior of high-B pulsars, such as PSR J1119-6127 which showed persistent abnormal radio pulsations in the months following its 2007 glitch \citep{weltevrede_glitch-induced_2011}. Note also that no significant radiative change had been associated with a glitch in a canonical radio pulsar until the dedicated observation of Vela in 2016 by \cite{palfreyman_alteration_2018}. 

This observation shows for the first time that the magnetosphere can be affected by a glitch -- an event considered to originate from the interior of the neutron star. We see no plausible mechanism for the coupling between the pulsar interior and the magnetosphere other than seismic motions of the crust (a quake). Excitation of seismic motions requires a sudden change of elastic stress on the timescale $\ll 1$ ms (the wave crossing time of the crust thickness). The quake is possible if the crust is stressed beyond its critical strain and ``fails'', launching shear waves. In this paper, we do not provide an argument for why a large stress should build up in Vela's crust. However, we argue that a quake is able to connect the 2016 glitch with the observed major magnetospheric disturbance coincident with the glitch.

The quake mechanism of exciting the magnetosphere of a neutron star was previously studied in several works [\cite{blaes_neutron_1989}; \cite{thompson_soft_1995}, \cite{timokhin_magnetosphere_2000}, \cite{timokhin_impact_2007}]. The wave transmission coefficient at the crust-magnetosphere interface was calculated by \cite{blaes_neutron_1989}, who considered quakes as possible triggers of gamma-ray bursts (GRBs). We consider much less energetic events, and thus we do not expect a bright GRB to accompany a glitch. Other key differences are that our model is two-dimensional (2D), time-dependent, and includes the self-consistent magnetic coupling to both the magnetosphere and the liquid core. These advances are essential for our model of the 2016 December event. We also include a liquid ocean, which was absent in the study of \cite{blaes_neutron_1989}, but find that it has little effect on the phenomena that we study. 

We find that the quake shear waves spread sideways and fill the whole crust. Therefore, seismic crustal oscillations populate the entire magnetosphere with Alfv\'en waves. The Alfv\'en waves bounce in the closed magnetosphere, become de-phased, and generate strong electric currents. De-phasing, in concert with growing wave amplitude in the outer magnetosphere leads to charge starvation, and $e^\pm$ discharge. The discharge can flood the magnetosphere with plasma, interrupting the observed radio emission. We also find that excitation of Alfv\'en waves in the liquid core efficiently drains energy from the crustal oscillations, and thus limits the quake duration. Assuming the mean magnetic field at the crust-core interface is comparable to the surface dipole field, and that the field in the core is bunched into flux tubes or domains (as is expected for type-II and type-I superconductors, respectively), we find that the quake amplitude is exponentially reduced on the timescale $\sim0.2$ s, fast enough to cause a single null.

 The paper is organized as follows. In Section \ref{vela_model} we present the relevant parameters of Vela, and other physics input required by our model. In Sections \ref{quake} and \ref{waves_and_discharge} we provide an analytic description of the proposed picture of the 2016 event. Section \ref{setup} outlines the formalism and numerical method for the full three-dimensional (3D) problem, although we only present results in 2D axisymmetry in this work. In Section \ref{results} we show four sample numerical models, and the results are further discussed in Section~\ref{discussion}.

\section{Vela model}\label{vela_model}

\subsection{Observed Parameters of the Vela pulsar}

The pulsar has spin period $P=2\pi/\Omega=89$ ms \citep{large_pulsar_1968}, and the light cylinder radius 
\beq
   R_\text{LC}=\frac{c}{\Omega}=4.2\times 10^8\,{\rm cm}.
\eeq 
Its spin-down rate $\dot{\Omega}=-9.8432\times 10^{-11}\text{ rad s}^{-2}$ gives a measurement of the magnetic dipole moment of the star $\mu_{\rm dip} = \sqrt{3 c^3 I \dot{\Omega}/(2\Omega^3)} \approx 3.4\times 10^{30} \text{ G cm}^3$, assuming $I\approx 10^{45}\text{ g cm}^2$ for the star's moment of inertia \citep{manchester_australia_2005}. The corresponding dipole magnetic field is $B_d\equiv \mu_{\rm dip}/r_\star^3=3.4\times10^{12}(r_\star/10\,{\rm km})^{-3}\,$G, where $r_\star$ is the neutron star radius. The spin-down power of Vela is given by 
\begin{equation}
\label{eq:Lsd}
   L_\text{sd}=I\Omega\dot{\Omega}\approx 7\times 10^{36} \text{ erg s}^{-1}.
\end{equation}
The pulsed radio emission at frequencies around 1.4~GHz has a much smaller luminosity \citep{manchester_australia_2005}, 
\begin{equation}
\label{eq:Lradio}
   L_\text{GHz} \approx 10^{28}\text{ erg s}^{-1}.
\end{equation}
The observed bolometric luminosity of the pulsar is dominated by GeV gamma-rays from the outer magnetosphere \citep{abdo_fermi_2009},
\begin{equation}
\label{eq:Lgamma}
  L_\text{GeV}\approx 8 \times 10^{34} \text{ erg s}^{-1}.
\end{equation}
The apparent surface temperature of Vela (as measured by a distant observer) is $T_\text{s}^{\infty} =(7.85\pm0.25)\times10^5$~K \citep{page_temperature_1996}. It is related to the actual surface temperature $T_\text{s}$ by $T_\text{s}^{\infty} = T_\text{s}  \sqrt{1-2GM/r_\star c^2}$ \citep{thorne_relativistic_1977}. We will use the approximate $T_\text{s} \approx 10^6\,$K.

\subsection{Magnetosphere, Ocean, Crust, and Core}

In the magnetosphere, the plasma mass density $\rho$ satisfies $\rho c^2\ll B^2/4\pi$, and so Alfv\'en waves propagate with almost the speed of light.
This changes in the ocean where density $\rho>\rho_B\equiv B^2/4\pi c^2$,
\begin{equation}
\rho_B =10^3\left(\frac{B}{3.4\times 10^{12}\,\rm G}\right)^2\text{ g cm}^{-3}.
\label{rho_B}
\end{equation}
The ocean is an excellent thermal conductor, and is effectively isothermal in the deeper layers. According to the temperature profiles of \cite{potekhin_atmospheres_2016} the ocean of a Vela-like pulsar with $T_\text{s} =10^6$ K has uniform temperature $T\sim10^8$ K for densities $\rho \gtrsim 10^6$ g cm$^{-3}$, which is in agreement with the analytic formula of \cite{gudmundsson_structure_1983}. The solid-liquid phase transition, which defines the top of the crust, is set
 by the Coulomb parameter $\Gamma=Z^2e^2/ak_B T \approx 175$, where $a=(4\pi n_i /3)^{-1/3}$ is the mean inter-ion spacing \citep{potekhin_equation_2000-1}. This defines the crystallization density
\begin{equation}
\rho_\text{crys} = 8\times 10^7  \left( \frac{T}{10^8}\right)^3\left(\frac{Z}{26}\right)^{-6}\left(\frac{A}{56}\right)\text{ g cm}^{-3},
\end{equation}
where $A$ and $Z$ are the ion mass and charge numbers. We adopt the value $\rho_\text{crys} = 10^8 \text{ g cm}^{-3}$ for all of our numerical simulations. \\

The density profile of the neutron star $\rho(r)$ (where $r$ is the radial coordinate) is obtained by integrating the equation of general relativistic hydrostatic equilibrium, using the SLy equation of state \citep{douchin_unified_2001}, with a central density $\rho=10^{15}\text{ g cm}^{-3}$. We use the OPAL equation of state for the ocean with temperature $T=10^8$ K \citep{rogers_opal_1996}. We also make use of the analytical fitting formula in \cite{haensel_analytical_2004} for the crust and the ocean. This gives a neutron star with mass $M=1.4M_\odot$ and radius $r_\star=11.69$ km. 

\begin{figure}[t]
\centering
\includegraphics[width=.46\textwidth]{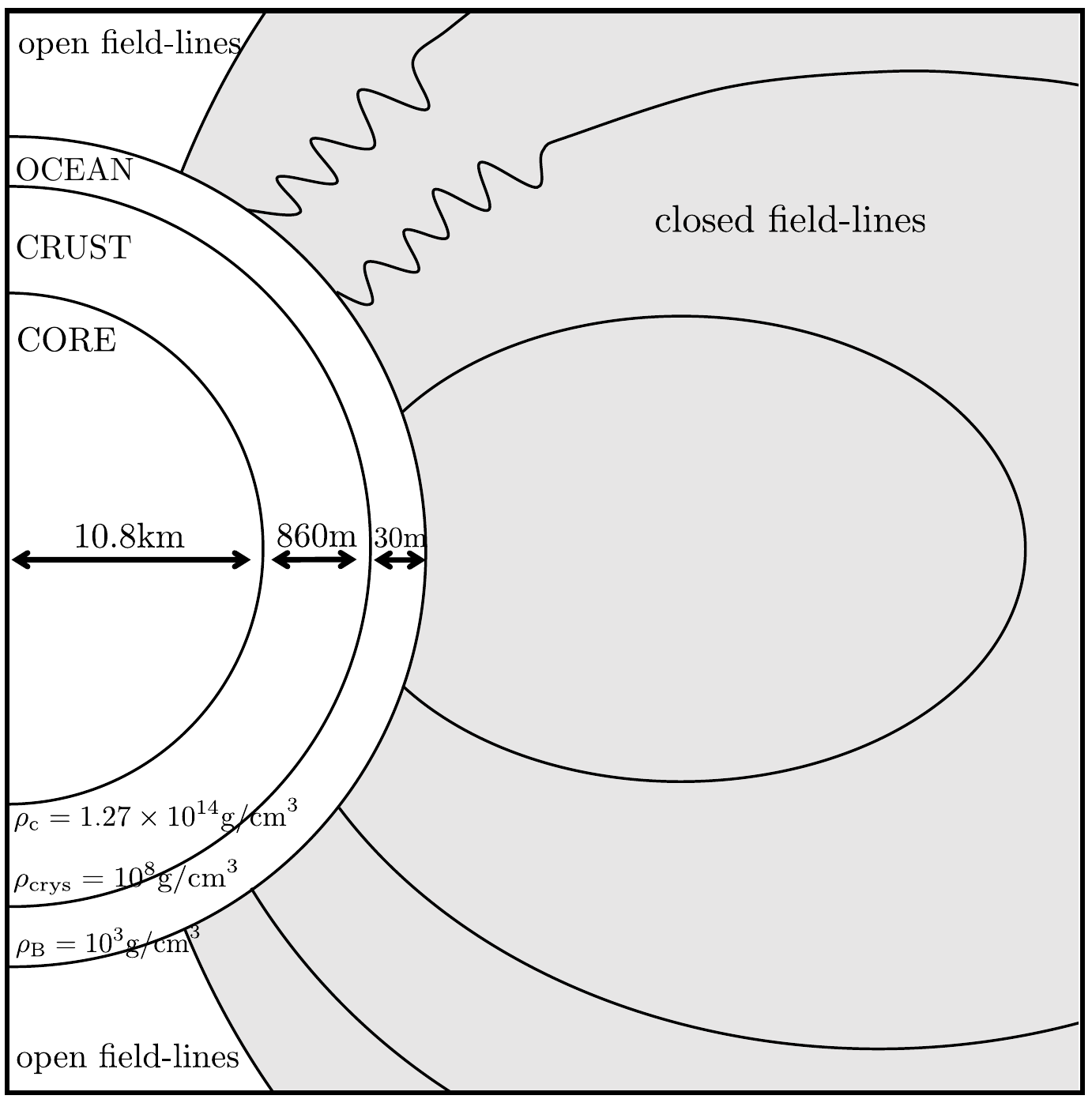}
\caption{Schematic picture of the neutron star and its magnetosphere, indicating relevant length scales and characteristic densities. The gray shaded region represents the closed magnetosphere.}
\label{ns_diagram}
\end{figure}

For the SLy equation of state, there is a phase transition at the bottom of the crust that occurs at fixed pressure $P=5.37\times10^{32} \text{ erg cm}^{-3}$. In our model, the crust-core boundary is located at $r_c=10.8$ km, with density
\begin{equation}
\rho_\text{c} = 1.27\times10^{14}\text{ g cm}^{-3}.
\end{equation}
The neutron star structure is summarized in Figure~\ref{ns_diagram}.
The crust-ocean boundary is located at radius $r_\text{crys}=11.66$ km, and
the thickness of the crust is $H\approx 860$ m. The mass of the crust is $M_c=1.6\times10^{-2}M_\odot$. The ocean is $\sim 30$ m deep. 

The speed of crustal shear waves is controlled by the shear modulus of the crustal lattice $\mu$. At densities far above the crystallization density, $\mu$ is proportional to the Coulomb energy density of the lattice and is approximately given by $\mu\approx 0.12\, n_i(Ze)^2/a$ where $a\sim n_i^{-1/3}$ is the separation of the ion lattice with density $n_i$ \citep{strohmayer_shear_1991}. At densities $\rho$ below the neutron drip density, $\rho_{\rm drip}\approx 4\times 10^{11}$~g~cm$^{-3}$, it gives $\mu\propto \rho^{4/3}$. In the deeper crust $\mu$ scales almost linearly with $\rho$. The shear modulus has a sharp cutoff at density $\rho_\text{crys}$, so that $\mu=0$ in the ocean. 

The star's magnetic field is frozen in its core, crust, and ocean. In our axisymmetric numerical models, we assume that the magnetic field in the magnetosphere has a dipolar configuration aligned with the axis of rotation. We also need to include magnetic stresses inside the crust, when computing the transmission of the seismic waves into the magnetosphere. For computational simplicity we assume that the field inside the crust is that of a monopole, chosen so that the field at the surface equals $ 3\times 10^{12}$ G. The spherical symmetry of the background configuration 
dramatically speeds up the computation of crustal oscillations, because the 
vibrational eigenfunctions used in our spectral code are easily computed through the separation of angular and radial variables (see Section 5.2 for details).\footnote{ Replacing the dipole field with monopole below the stellar surface only slightly changes the crust dynamics and the calculated displacements of the magnetospheric footpoints. In the magnetosphere itself, the waves are followed in the correct dipole background. Had we kept the dipole field throughout, we would get similar results with a much greater computational
effort.
}
An important feature of our model is that the magnetic field lines connecting the rotating star with the light cylinder are assumed to be open, and their footprints on the star form the two ``polar caps.'' In the simplest case of a nearly aligned rotator, the angular  size of the polar cap is $\theta_p\approx(r_\star/R_\text{LC})^{1/2}\approx 0.05$. 

\medskip


\section{Quake excitation of shear waves}
\label{quake}

We model the quake as a sudden change in shear stress in the deep crust, which launches an elastic wave with an initial strain amplitude $\epsilon_0$. The quake is triggered in a region of vertical thickness $\Delta \ell \sim 10^4$~cm (comparable to the hydrostatic pressure scale height) and horizontal area $\AQ$. The energy of the quake is 
\begin{equation}
\label{eq:EQ}
E_Q\sim \frac{\mu \epsilon_0^2}{2}\,\Delta \ell \AQ \sim 10^{39} \left(\frac{\epsilon_0}{10^{-3}}\right)^2 
\left(\frac{\AQ}{10^{11}\,{\rm cm}^2}\right) \text{ erg}.
\end{equation}
The wave propagates toward the stellar surface with speed $v_s=(\mu/\rho)^{1/2}\approx 10^8$~cm~s$^{-1}$ and crosses the crust thickness $H\sim 10^5\,$cm on the timescale
\begin{equation}
   \tauc\sim \frac{H}{v_s} \sim 1\text{ ms}.
\end{equation}
The thickness of the shear layer sets the characteristic frequency of the generated waves. 
As a concrete example, consider the smooth deformation 
\beq
\xi(z) =\frac{\xi_0}{2}\text{erf}\left[\frac{\sqrt{2}(z-z_Q)}{\Delta\ell}\right],
\eeq
where $z<0$ is the distance below the stellar surface. It corresponds to a shear layer of thickness $\Delta\ell$ at depth $z_Q$.  The characteristic length scale of the deformation is $\lQ\equiv \xi(d\xi/dz)^{-1} = \sqrt{\pi/8}\Delta\ell$. The characteristic angular frequency is

\begin{equation}
 \omega\sim \frac{v_s}{\lQ}\approx 2\times 10^4 \left(\frac{\Delta\ell}{10^4\,\rm cm}\right)^{-1} {\rm rad~s}^{-1}.
\end{equation}
The quake can excite a broad spectrum of waves extending to frequencies well above this characteristic frequency.
\medskip

\subsection{One-dimensional Model of Waves}

Much insight about the transmission of seismic
waves into the magnetosphere and the core  can be obtained from studying the propagation and transmission of radially directed seismic waves. A classic one-dimensional (1D) model of this type was developed by \cite{blaes_neutron_1989}. Following their approach, we approximated the crust as a 1D slab with the normal along the $z$-axis (which would be in the radial direction for a spherical crust). The shear displacement $\xi(z)$ is in the $\hat{y}$-direction. For the timescales of interest, the star is an ideal conductor, so the magnetic field is perturbed by the displacement along the $y$-axis, $B_y = B_z\partial \xi  / \partial z$, as required by the flux-freezing condition. As a first approximation, the magnetosphere is also described by ideal MHD.

\begin{figure}[t]
\centering
\includegraphics[width=.46\textwidth]{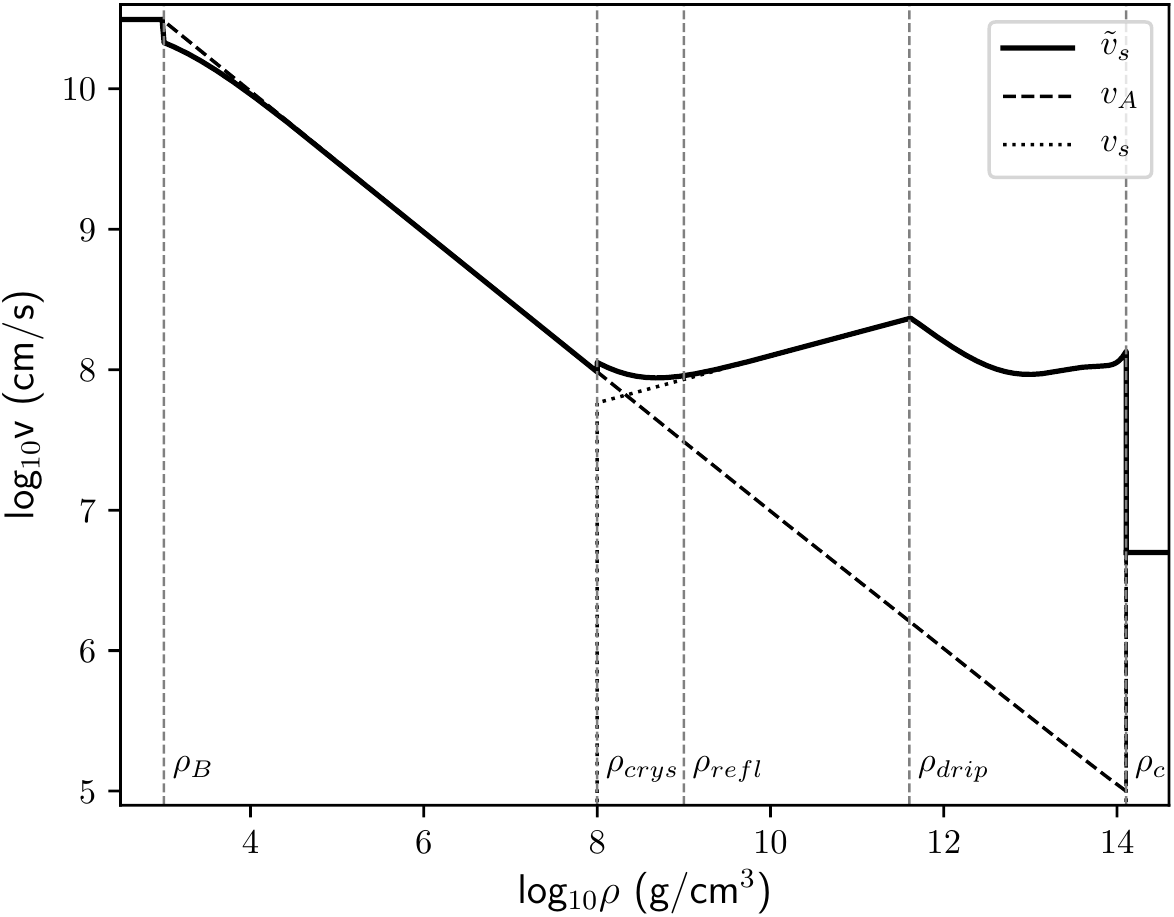}
\caption{Wave speed $\tilde{v}_s(\rho)$ in the magnetosphere, ocean, crust, and core (thick black line). The dashed line shows the Alf\'en speed $v_A(\rho)$, and the dotted line shows the elastic wave speed $v_s(\rho)$.}
\label{velocity_profile}
\end{figure}

The magneto-elastic wave equation is given by
\begin{equation}
\tilde{\rho}\,\frac{\partial^2\xi}{\partial^2 t} = \frac{\partial}{\partial z}\left( \tilde{\mu}\frac{\partial\xi}{\partial z}\right),
\label{1d_wave}
\end{equation}
where $\tilde{\rho}$ and $\tilde{\mu}$ are the effective mass density and shear modulus, respectively, such that
\begin{align}
\tilde{\rho} = \rho + \frac{B_z^2}{4\pi c^2}, &&\tilde{\mu} = \mu + \frac{B_z^2}{4\pi}.
\label{mu_ef}
\end{align}
The wave speed is given by $\tilde{v}_s=(\tilde{\mu}/\tilde{\rho})^{1/2}$ and shown in Figure~\ref{velocity_profile}. It equals $v_s\approx 10^8$~cm~s$^{-1}$ in the deep crust and grows to the speed of light in the magnetosphere. The wave speed in the liquid core equals the Alfv\'en speed, which depends on $B$ and the density of matter coupled to the Alfv\'en wave, as discussed in Section~\ref{core} below.

For a harmonic time dependence $\xi\propto e^{-i\omega t}$ with $\omega\gtrsim 10^4$~rad~s$^{-1}$ the wave propagation may be described in the WKB approximation. Then an upward propagating wave and its reflection from the low-density surface layers are given by \cite{blaes_neutron_1989}
\begin{equation}
\xi 
\propto \frac{1}{\sqrt{\rho \tilde{v}_s}}
\left[ e^{-i(u + \omega t)} + A_R e^{i(u-\omega t)}\right],
\label{WKB}
\end{equation}
where 
\begin{equation}
u\equiv -\int^z dz' \frac{\omega}{\tilde{v}_s}.
\end{equation}
The first term in brackets in Equation \eqref{WKB} is the upward propagating wave, and the second term 
with the
complex amplitude $A_R$ is the reflected wave. The scaling of the overall amplitude $\xi\propto(\rho\tilde{v}_s)^{-1/2}$ comes from the conservation of energy flux in the wave $F\sim \rho v_s \omega^2\xi^2$.
In particular, using $\tilde{v}_s\propto \rho^{1/6}$ in the upper crust, one finds
\begin{equation}
   \xi\propto\rho^{-7/12} \qquad (\rho<\rho_{\rm drip}),
\end{equation}
and the strain in the shear wave is 
\begin{equation}
  \epsilon\equiv \frac{\partial\xi}{\partial z}=\frac{\xi\omega}{\tilde{v}_s}\propto \frac{1}{\rho^{1/2}\, \tilde{v}_s^{3/2}}\propto \rho^{-3/4}.
\end{equation}
The strain can become large in the low-density regions and cause a secondary failure of the crust. However, in this work we choose to remain within the linear theory of elasticity, which is applicable in the limit of $\epsilon\ll 1$. In particular we assume that nowhere in the solid crust does the strain exceed the critical value $\epsilon_\text{crit}\sim0.1$ \citep{horowitz_breaking_2009}. This condition is satisfied for a quake with a typical strain in the deep trigger region $\epsilon_0< 2\times 10^{-3}$. Our  numerical models in Section~\ref{results} have the starquake area $\AQ\sim 3\times 10^{11}\,$cm$^2$, which gives the quake energy $E_Q\sim 10^{38}$~erg (Equation~\ref{eq:EQ}).

For waves excited on scales comparable to the hydrostatic scale height of the crust 
(as assumed in our quake scenario), the WKB approximation is not accurate, and the exact solution should be obtained numerically. More importantly, the 1D  model is insufficient, as the quake waves propagate at different angles, and after reflection from the surface layers, they tend to spread sideways to fill the entire crust. The numerical simulations of this process are presented in Sections~\ref{setup} and \ref{results} below. Here we estimate the transmission coefficients analytically using the simple 1D model.
\medskip

\subsection{One-dimensional Wave Transmission into the Magnetosphere}
The wave reflection occurs in the upper crust, which is defined by $\rho_{\rm drip}<\rho<\rho_{\rm crys}$. For a vertically propagating wave, the transmission coefficient (the ratio of transmitted to incident energy flux) is given by 
\beq
  \T_m = \frac{4Z_\text{crust}Z_\text{mag}}{(Z_\text{crust}+Z_\text{mag})^2}
 \approx \frac{4Z_\text{mag}}{Z_\text{crust}},
\eeq
where the impedance $Z=\tvs\tilde{\rho}$ is evaluated in the upper crust at the transmission layer $\rho_\text{refl}$,  $Z_{\rm crust}\approx \rho_{\rm refl}\,\tvs(\rho_{\rm refl})$, and in the magnetosphere, $Z_{\rm mag}\approx \rho_B c\ll Z_{\rm crust}$. In the relevant region, $\tilde{\rho}\approx\rho\gg \rho_B$,  and the shear wave speed may be approximated as
\beq 
\label{eq:vsappr}
    \tilde{v}_s^2 \approx 10^{15} \left( 7\rho_9^{1/3} + \frac{b^2}{\rho_9}\right)\frac{{\rm cm}^2}{{\rm s}^2}, 
    \quad b=\frac{B}{3.4\times 10^{12}\,\rm G},
\eeq
where we normalized $B$ to the characteristic dipole field of the Vela pulsar, and $\rho_9 = \rho / (10^9$~g~cm$^{-3})$. Note that $\tvs(\rho)$ is non-monotonic (see Figure~\ref{velocity_profile}). The wave speed first decreases from $v_s\approx 2\times 10^8$~cm~s$^{-1}$ in the deep crust to $9\times 10^7$~cm~s$^{-1}$ at $\rho=10^9$~g~cm$^{-3}$. This decrease shortens the wavelength by a factor of $\sim 2$, so that it remains comparable to or shorter than the hydrostatic scale height.
However, as $\rho$ further decreases below $10^9$~g~cm$^{-3}$, the wave speed steeply grows, and the length scale of this change soon becomes shorter than the wavelength. Therefore, reflection mainly occurs at $\rho_{\rm refl}$ just below $10^9$~g~cm$^{-3}$. The reflection condition may be written as \citep{blaes_neutron_1989}
\beq
\label{eq:refl1}
     \left|\frac{d}{dz} \tilde{v}_s^2\right| \sim \omega \tilde{v}_s.
\eeq
Pressure in the upper crust is dominated by relativistic degenerate electrons, and the hydrostatic balance gives the relation $|z|\approx 6\times 10^3 \rho_9^{1/3}$~cm, where $z<0$ is the depth below the stellar surface. Using this relation and Equation~(\ref{eq:vsappr}), we obtain the equation for $\rho_{\rm refl}$,
\beq
\label{eq:refl}
  \left| 1 - \frac{3b^2}{7\rho_9^{4/3}}\right| \sim \frac{\omega_4}{\sqrt{2}} \left( \rho_9^{1/3} + \frac{b^2}{7\rho_9}\right)^{1/2}.
\eeq
At high frequencies one can keep only the second terms on both sides of the equation, which gives 
\beq
\label{eq:rho_refl}
   \rho_{\rm refl}\approx 2 \times 10^9 \left(\frac{b}{\omega_4}\right)^{6/5} \frac{\rm g}{{\rm cm}^3} 
    \qquad \left(\omega> \omega_\text{eva} \right).
\eeq  
One can show that the reflection condition~(\ref{eq:refl1}) does not apply when $\omega<\omega_{\rm eva}\approx 2\times 10^4$~rad~s$^{-1}$. In that case, the reflection occurs deeper in the crust  due to the appearance of an evanescent zone, and the transmission coefficient becomes suppressed as $(\omega/\omega_{\rm eva})^7$ [see \cite{blaes_neutron_1989}]. Note also that at frequencies $\omega\lesssim v_s/H\approx 10^3$~rad~s$^{-1}$ the crust oscillates as a whole and directly moves the footprints of the magnetospheric field lines\footnote{The fundamental frequency of the liquid ocean is $\omega_\text{ocean} = \tilde{v}_s/H_o \sim 3 \times 10^5\text{ rad s}^{-1}$ where $H_o\sim 30$~m is the scale height of the ocean. For the characteristic frequency of the crustal oscillations $\omega\ll\omega_\text{ocean}$, the ocean can be viewed as attached to the moving crust. Effectively, the waves are transmitted directly from the solid crust to the extended magnetosphere above the ocean.} . 

Using Equation~(\ref{eq:rho_refl}) for $\rho_{\rm refl}$ and the corresponding $\tvs(\rho_{\rm refl})\approx  3\times 10^7\,b\,\rho_9^{-1/2}$~cm~s$^{-1}$, we find\footnote{\cite{blaes_neutron_1989} obtained a different result $\T_m\propto B^{4/7} \omega^{3/7}$,  because they considered neutron stars with lower $B=10^{11}\,$G. In that case, $\rho_{\rm refl}$ is much lower, and the hydrostatic stratification is different because the degenerate electrons are sub-relativistic.}
\beq
   \T_m\approx
\begin{cases}
   3\times 10^{-3}\,b^{2/5}\,\omega_4^{3/5} &\qquad \left(  \omega_\text{eva} < \omega< \omega_\text{crys} \right) \\
    10^{-2}\, b^{2} &\qquad \left(  \omega \geq \omega_\text{crys} \right),
\end{cases}
\label{eq:T_m}
\eeq
where $\omega_\text{crys} \approx 10^5$~rad~s$^{-1}$ is the frequency at which the waves reflect near the crust-ocean interface, $\rho_\text{refl}(\omega_\text{crys})=\rho_\text{crys}$. All waves with frequency $\omega \geq \omega_\text{crys}$ experience substantial reflection at the solid-liquid phase boundary (note the discontinuity in $\tvs(\rho_\text{crys})$ in Figure \ref{velocity_profile}). The frequency independence of $\T_m$ at $\omega>\omega_\text{crys}$ was not seen in \cite{blaes_neutron_1989} because they did not include the sharp phase transition at the top of the crust.

A large fraction of the quake energy is deposited into waves with $\omega\gtrsim 2\times 10^4$~rad~s$^{-1}$, and these waves will leak into the magnetosphere with the above transmission coefficient. 
\medskip

\subsection{Wave Transmission into the Core}
\label{core}
The bottom of the crust is magnetically coupled to the liquid core. The core supports a multitude of MHD modes, which get excited while draining elastic wave energy from the crust \citep{levin_qpos_2006}. The Alfv\'en crossing time of the core $\tau_A \sim r_\star/v_A \sim 1\text{ s}$ is longer than the characteristic lifetime of crustal waves (estimated below). Effectively, the waves escape into the core as if it were an infinite reservoir. Under such conditions, the transmission coefficient for a vertically propagating shear wave at the crust-core interface can be estimated as
\begin{equation}
 \T_c = \frac{4Z_\text{crust}Z_\text{core}}{(Z_\text{crust}+Z_\text{core})^2},
\end{equation}
where $Z_\text{crust}$ and $Z_\text{core}$ are the
impedances 
of the crust and the outer core,
\begin{align}
Z_\text{crust} = \rho_> \tilde{v}_s, && Z_\text{core} = \rho_< v_A.
\end{align}
Here $\rho_>$ and $\rho_<$ are the mass densities of the matter that participate in the oscillations above and below the crust-core interface, respectively.

For typical pulsar parameters $Z_\text{crust}\gg Z_\text{core}$, and the transmission coefficient is
\begin{equation}
 \T_c \simeq \frac{4 Z_\text{core}}{Z_\text{crust}} = 4\frac{\rho_<}{\rho_>}\frac{v_A}{\tilde{v}_s}.
\end{equation}
In the deep crust (below the neutron drip), a large fraction of mass is carried by free superfluid neutrons. However, entrainment is probably very strong, and we assume that free neutrons couple to shear waves, so that $\rho_>$ equals the total local density of the crust $\rho$ \citep{carter_entrainment_2006}.

By contrast, in the superfluid core neutrons become decoupled from the oscillations. Furthermore, as long as protons are superconducting, the magnetic flux is bunched into flux tubes with field $B_c\sim 10^{15}$~G.
This causes two effects of superfluidity and superconductivity
on wave transmission into the core:
\\
i) The effective tension of magnetic field lines in the core is $BB_c/4\pi$. Therefore, bunching of the magnetic field into quantized flux tubes dramatically increases the magnetic tension, by a factor of $B_c/B\sim 300$. This enhances the transmission coefficient by a factor of $\sim 20$.
\\
ii) Decoupling of protons from other species in the core reduces the effective mass density participating in the oscillation to the proton density, $\rho_< = \rho_p$.\footnote{Even in the presence of strong vortex-flux-tube interactions, a negligible fraction of the neutron mass couples to the oscillations we are considering [see \cite{van_hoven_hydromagnetic_2008}]}
This reduction of $\rho_<$ (by a factor of $\sim 10$) decreases the transmission coefficient by a factor $\sim 3$.
\\
The net effect is an enhancement of the transmission coefficient $\T_c$, by a factor of $\sim 6$.

The Alfv\'en speed in the outer core is
\beq
 v_A=\left( \frac{B B_{c}}{4\pi\rho_p}\right)^{1/2}
      \sim 5\times 10^{6}\,\text{cm s}^{-1},
\eeq
and the resulting transmission coefficient is
\beq
   \T_c\sim2\times 10^{-2}.
\eeq
The transmitted waves are lost for the quake. Since $\T_c$ for the superconducting core is $\sim 5$ times greater than $\T_m$, the lifetime of crustal waves is controlled by their leakage to the core rather than to the magnetosphere. 
The characteristic lifetime is given by 
\beq
   \tau_\text{core}=\frac{2\tauc}{\T_c}\sim 100\,{\rm ms}.
\eeq

\medskip

\section{Magnetospheric waves and electric discharge}
\label{waves_and_discharge}

\subsection{Electric Current of Alfv\'en Waves}

The magnetospheric disturbance may be described as ideal MHD Alfv\'en waves as long as there is enough plasma in the magnetosphere to support electric currents. The energy flux of the Alfv\'en waves into the magnetosphere is approximately given by
\begin{equation}
   \FA_\star \sim \frac{E_Q\T_{\rm m}}{ \tauc A} \sim 4\times 10^{26}\, 
   \frac{E_{Q,38}}{A_{12}}\, \frac{\rm erg}{{\rm s\,cm}^2},
\end{equation}
where $A$ is the area through which the crustal wave energy is leaking into the magnetosphere. Initially, at times comparable to $\tauc=H/v_s\sim 1\,$ms, the waves emerge from the quake area $A\approx \AQ$. Later, $A$ grows as the waves spread horizontally through the crust. 

The Alfv\'en waves are ducted along the magnetic field lines, and their flux $\FA$ changes proportionally to the local magnetic field $B$, 
\beq
\label{eq:FA}
 \FA=\FA_\star\, \frac{B}{B_\star}.
\eeq
This fact follows from $F\,dS=$const where $dS=d\psi/B$ is the cross-sectional area of a field-line bundle carrying infinitesimal magnetic flux $d\psi$. The flux $\FA$ determines the wave amplitude $\delta B$,
\begin{equation}
 \delta B \approx \left(\frac{8\pi \FA}{c} \right)^{1/2} \sim 3 \times 10^8 \, \FA_{26}\text{ G}.
\end{equation}
The relative perturbation of the magnetic field is small near the star,
$\delta B_\star/B_\star \approx 10^{-4}\,F_{\star,26}^{1/2}$.
However, it grows for waves propagating to radii $r\gg r_\star$ in the outer magnetosphere as $\delta B/B\propto F^{1/2}/B\propto B^{-1/2}$. In particular, for a dipole magnetosphere, $B\propto r^{-3}$, and so
\beq
\label{eq:ampl}
 \frac{\delta B}{B} \approx 10^{-4}\,F_{\star,26}^{1/2} 
 \left(\frac{r}{r_\star}\right)^{3/2}.
\eeq
The emitted Alfv\'en waves bounce in the closed magnetosphere on the light-crossing  timescale $t_b$ and can accumulate energy and $\delta B$ during the quake. This accumulation occurs on field lines that do not extend too far from the star, so that their $t_b$ is shorter than the quake duration.

Alfv\'en waves can be thought of as the propagating shear of the magnetic field lines. They require electric current $j_\parallel$ along $\vec{B}$ as long as the wavevector $\vec{k}$ has a component perpendicular to $\vec{B}$, $k_\perp\neq 0$. This component is inevitably present, since the field lines are curved. The waves 
develop different phases on different field lines, and thus amplify the gradients of $\delta{B}$ in the direction perpendicular to 
the field lines.

The electric current $j_\parallel$ may be estimated as
\footnote{In particular, in axisymmetry, $\delta \vec{B}$ is azimuthal, and its gradient is in the poloidal plane. This gradient has a component perpendicular to the background dipole field $\vec{B}$ and generates $\nabla\times\vec{\delta B}\parallel\vec{B}$.}
\beq
\label{eq:jpar}
   j_\parallel \sim \frac{c}{4\pi}\,k_\perp \delta B \sim
     \frac{c}{4\pi}\,\frac{\delta B}{\lperp},
\eeq
where $\lperp\sim k_\perp^{-1}$ is the spatial scale of the wave variation perpendicular to $\vec{B}$.
The length scale $\lperp$ is initially determined by the elasto-dynamics of the crust. But once Alfv\'en waves on neighbouring field lines accumulate a difference in path length similar to the wavelength, they are effectively de-phased. Therefore, $\lperp$ decreases, and so $j_\parallel$ grows as the Alfv\'en waves keep bouncing in the closed magnetosphere. The growth of $j_\parallel$ may be estimated as follows.\\
 
Let us consider a dipole magnetosphere and let $\theta$ be the polar angle measured from the dipole axis. It is convenient to label the field lines by the poloidal magnetic flux function,
\beq 
\label{eq:psi}
   \psi=\frac{\mu_{\rm dip} \sin^2\theta}{r}, 
\eeq
which is constant along a field line. In the axisymmetric magnetosphere, $\psi=$const on each flux surface formed by a field line rotated about the axis of symmetry. A closed field line with footprints on the star at $\theta_\star$ and $\pi-\theta_\star$ extends to radius $r_{\max}=r_\star/\sin^2\theta_\star$, and its length is $\sim 3r_{\max}$. The bounce cycle of Alfv\'en waves along a closed field line takes time $t_b\propto r_{\rm max}\propto \psi^{-1}$, so two field lines separated by a small $\Delta\psi$ have different $t_b$,
\begin{equation}
    \frac{\Delta t_b}{t_b} \approx -\frac{\Delta\psi}{\psi}.
\end{equation}
After time $t$, the accumulated phase mismatch between waves on flux surfaces separated by $\Delta \psi$ is 
\beq
  \frac{\Delta\phi}{\omega t} \approx -\frac{\Delta\psi}{\psi}.
\eeq
De-phasing on a given scale $\Delta\psi_{\rm de}$ occurs when $|\Delta\phi| \sim \pi$, and so $\Delta\psi_{\rm de}(t)\sim \pi \psi/\omega t$.  At a radius $r>r_\star$, the distance $\lperp$ between the poloidal field lines separated by $ \Delta\psi_{\rm de}$ is 
\beq
   \lperp(t) \approx r\,\frac{\Delta\psi_{\rm de}}{\partial \psi/\partial \theta}
   \sim \frac{\pi r \tan\theta}{2\omega t}.
\eeq
This gives the current density (Equation~\ref{eq:jpar})
\beq
\label{eq:jpar1}
   j_\parallel (t)\sim  \frac{c\,\delta B}{2\pi^2r \tan\theta}\;\omega t.
\eeq
\medskip

\subsection{$e^\pm$ Discharge}
\label{discharge}
In the canonical pulsar picture, the rotating closed magnetosphere is filled with plasma that sustains the corotation electric
 field $\vec{E}=-\vec{v}\times\vec{B}/c$ (here $\vec{v}=\vec{\Omega}\times\vec{r}$). This implies the characteristic minimum plasma density \citep{goldreich_pulsar_1969},
\beq
  n_{\rm GJ}=\frac{|\nabla\cdot\vec{E}|}{4\pi e}
  \approx  \frac{|\vec{\Omega}\cdot\vec{B}|}{2\pi ce}.
\eeq
The actual plasma density may be higher by a multiplicity factor $\M$, $n=\M n_{\rm GJ}$. 
This factor is believed to be large in the open field-line bundle, in some cases exceeding $10^3$, because the open field lines are twisted and sustain continual $e^\pm$ discharge. The value of $\M$ in the closed magnetosphere is unknown and likely much lower, because this zone is not active and generates no discharge. It may, however contain $e^\pm$ pairs created by gamma-rays entering from the open field lines \citep{chen_electrodynamics_2014}.

The existing plasma in the closed zone can sustain Alfv\'en waves with the maximal current 
\beq
\label{eq:jmax}
   j_{\max}=ce\M n_\text{GJ}=\frac{\M\, |\vec{\Omega}\cdot\vec{B}|}{2\pi}.
\eeq
When $j_\parallel$ exceeds $j_{\max}$, the waves become charge starved, and the ideal MHD approximation must break \citep{blaes_neutron_1989}. From Equations~(\ref{eq:ampl}), (\ref{eq:jpar1}), and (\ref{eq:jmax}), we find
\begin{eqnarray}
\nonumber
  \frac{j_\parallel}{j_{\max}}
  & \sim & \frac{c\,(\delta B/B)\,\omega t}{4\pi^2\M\,\Omega\,r \tan\theta} \\
  & \sim & 10\, \frac{\omega_4}{\M\tan\theta}  
    \left(\frac{\delta B_\star/B_\star}{10^{-4}}\right)
    \left(\frac{r}{r_\star}\right)^{1/2}
    \left(\frac{t}{0.1\,\rm s}\right).
\end{eqnarray}
One can see that the Alfv\'en waves generated by the quake can become charge-starved, especially when one takes into account the growth of $\delta B_\star$ due to the accumulation of waves trapped in the closed magnetosphere.\\

Once charge starvation is reached, a parallel electric field will be induced to support $\nabla \times \vec{B}$. The resulting parallel voltage may be estimated as
\begin{equation}
\Phi \sim \frac{4\pi j_\parallel}{c} \lperp^2 \sim \delta B \,\lperp.
\label{voltage}
\end{equation}
The voltage is maximum for the largest $\lperp$ at which starvation occurs. This scale $\lperp$ is given by the condition
\begin{equation}
   \frac{\delta B}{\lperp} \sim 4\pi \M \rho_\text{GJ},
\end{equation}
which yields
\begin{equation}
 \Phi \sim \frac{c (\delta B)^2}{2 \M \Omega B}
 =\frac{4\pi F}{\M \Omega B}.
\end{equation}
Note that $F/B=$const (Equation~\ref{eq:FA}), so the generated voltage is approximately the same at all $r$ along the field line and can be estimated with $F=F_\star$ and $B=B_\star$. This gives
\beq
   \frac{e\Phi}{m_e c^2}\sim 3\times 10^9 \,\M^{-1} \, F_{\star,26}.
\eeq
This voltage exceeds the threshold for $e^\pm$ discharge, as particle acceleration to $\gamma\sim 10^6-10^7$ is sufficient to ignite $e^\pm$ creation by emitting high-energy curvature photons \citep{ruderman_theory_1975}. This process will flood the magnetosphere and the open field-line bundle with $e^\pm$ plasma. Therefore, the quake should be capable of interrupting the normal radio pulsations of Vela.
\medskip

\section{Setup of the Numerical Simulation}\label{setup}

In this section, we outline the formalism and the setup of our numerical simulations. We are able to simulate the elasto-dynamics of the crust in 3D; however we are currently limited to the 2D axisymmetric simulations of the magnetosphere. Since the two computations are coupled, we are restricting ourselves to the 2D axisymmetric simulations of the whole system.

\medskip
\subsection{Dynamics of the Crust}

We use the linearized equations of motion (see, e.g. \cite{mcdermott_nonradial_1988}, \cite{blaes_neutron_1989}). For simplicity, the background state of the crust is assumed to have a potential magnetic field, $\nabla\times\vec{B}=0$ and $\vec{j}=0$. The background is static and has $\vec{E}=0$. A displacement $\vec{\xi}(t,\vec{r})$ creates motion with velocity $\dot{\vec{\xi}}=d\vec{\xi}/dt\approx \partial \vec{\xi}/\partial t$ in the linear order. The momentum and continuity equations are
\begin{equation}
 \rho\,\ddot{\vec{\xi}}
= \nabla\cdot\vec{\sigma} + \frac{1}{c} \delta\vec{j}\times\vec{B} + \vec{g}\,  \delta\rho - \nabla\delta p,
\label{momentum}
\end{equation}
\begin{equation}
\delta\rho = -\nabla\cdot(\rho\,\vec{\xi}),
\label{mass}
\end{equation}
where $\vec{\sigma}$ is the elastic stress tensor of the crustal Coulomb lattice, $\vec{g}$ is the gravitational acceleration, and $p$ is the pressure; perturbations are denoted by $\delta$. The quake waves involve a fraction of the Coulomb energy density of the lattice, which is much smaller than the hydrostatic pressure. Therefore, compressive motions and radial displacements are negligible, and hereafter we consider only solenoidal deformations ($\nabla\cdot\vec{\xi}=0$) and set $\xi_r=0$. In this model, $\delta\rho=0$, $\delta p=0$, and the density of the crust is spherically symmetric. 

The stress tensor for an isotropic and incompressible solid is \citep{landau_theory_1970},  
\begin{equation}
\sigma_{ij} = \mu\left(\frac{\partial\xi_i}{\partial x_j} + \frac{\partial\xi_j}{\partial x_i} \right), 
\label{stress}
\end{equation}
where $\mu$ is the crustal shear modulus. The linear theory of elasticity is applicable in the limit of small strain. 

For the short timescales considered in this problem, the crust is effectively an ideal conductor. In the conductor rest frame, which is moving with velocity $\dot{\vec{\xi}}$, the electric field must vanish,
\beq
  \delta\vec{E}+\frac{\dot{\vec{\xi}}\times\vec{B}}{c}=0. 
\eeq
Then the induction equation $\partial \vec{B}/\partial t=-c\nabla\times\vec{E}$ gives
\beq
    \delta\vec{B}=\nabla\times (\vec{\xi}\times \vec{B}).
\eeq
The excited electric current $\delta\vec{j}$ is related to $\delta\vec{B}$ and $\delta\vec{E}$ by the Maxwell equation,
\begin{eqnarray}
\nonumber
  \frac{4\pi}{c}\delta\vec{j} 
  & = & \nabla \times \delta \vec{B} - \frac{1}{c}\,\frac{\partial \delta \vec{E}}{\partial t}  \\
  & = & \nabla \times \nabla\times (\vec{\xi}\times \vec{B})  +\frac{1}{c^2}\,\ddot{\vec{\xi}}\times\vec{B}. 
\label{eq:dj}
\end{eqnarray}

Substitution of Equations~\eqref{stress} and \eqref{eq:dj} into Equation \eqref{momentum} gives the elasto-dynamic wave equation,
\begin{equation}
\begin{aligned}
 \rho\,\ddot{\vec{\xi}}+\rho_B\,\ddot{\vec{\xi}}_\perp
  = &(\nabla\mu \cdot\nabla)\vec{\xi} - (\vec{\xi}\cdot \nabla)\nabla\mu  + \mu\nabla^2\vec{\xi} \\
+ & \frac{1}{4\pi}[\nabla \times \nabla \times (\vec{\xi}\times\vec{B})]\times \vec{B},
\end{aligned}
\label{wave_eqn}
\end{equation} 
where $\rho_B=B^2/4\pi c^2$ and $\vec{\xi}_\perp$ is the displacement perpendicular to $\vec{B}$. In the crust, Equation \eqref{wave_eqn} describes oscillations of the magnetized solid. In the liquid ocean, $\mu\longrightarrow 0$ and Equation \eqref{wave_eqn} describes pure Alfv\'en waves. The dynamics of the crust and the ocean of interest occurs at densities $\rho\gg\rho_B\sim 10^3$~g~cm$^{-3}$ where the term $\rho_B\,\ddot{\vec{\xi}}_\perp$ can be neglected.
\medskip

\subsection{Spectral Method} \label{spectral_method}

In order to numerically solve Equation \eqref{wave_eqn}, we prefer to use a spectral method for superior stability and accuracy over a large range of densities. Our formalism follows closely that of \cite{van_hoven_magnetar_2012}. Equation \eqref{wave_eqn} is written in the form
\begin{equation}
\frac{\partial^2\vec{\xi}}{\partial t^2} = \mathcal{\hat{L}}(\vec{\xi})
  =\mathcal{\hat{L}}_{\text{el}}(\vec{\xi}) + \mathcal{\hat{L}}_{\text{mag}}(\vec{\xi}),
\label{xieqn}
\end{equation}
where the linear differential operators $\mathcal{\hat{L}}_{\text{el}}$ and $\mathcal{\hat{L}}_{\text{mag}}$ give the acceleration due to elastic and magnetic forces, respectively. The elastic acceleration is
\begin{equation}
\mathcal{\hat{L}}_{\text{el}}(\vec{\xi}) = \frac{1}{\rho}\big[(\nabla\mu \cdot\nabla)\vec{\xi} - (\vec{\xi}\cdot \nabla)\nabla\mu  + \mu\nabla^2\vec{\xi}\big].
\label{L_el}
\end{equation}
The operator $\mathcal{\hat{L}}_{\text{mag}}$ is greatly simplified by approximating the crustal magnetic field as purely radial (a monopole) with $B_r = B_0(r_\star/r)^2$, where $B_0$ is the typical magnetic field strength in the crust. In reality $B_r$ varies over the crust. We use the fiducial value of $B_0=3\times10^{12}$~G. The magnetic acceleration is then
\begin{equation}
\mathcal{\hat{L}}_{\text{mag}} =\frac{1}{4\pi\rho}[\nabla \times \nabla \times (\vec{\xi}\times\vec{B})]\times \vec{B} = r\frac{\mu_B}{\rho} \frac{\partial^2}{\partial r^2}\left(\frac{\vec{\xi}}{r} \right),
\label{L_mag}
\end{equation}
where $\mu_B \equiv B_r^2/4\pi$ depends only on $r$. We use spherical coordinates $r,\theta,$ and $\phi$.

We separate variables $t,r,\theta,$ and $\phi$ in Equation~(\ref{xieqn}), and define magneto-elastic modes $\vec{\xi}_{nlm}$ as the eigenfunctions of the operator $\mathcal{\hat{L}}$ with the boundary conditions of zero stress at the boundaries (free oscillations of the system),
\begin{equation}
\mathcal{\hat{L}}(\vec{\xi}_{nlm}) = -\omega_{nlm}^2\vec{\xi}_{nlm}.
\end{equation}
Here $\omega_{nlm}$ is the eigenfrequency of the mode with radial, polar, and azimuthal numbers $n$, $l$, and $m$, respectively. The modes $\vec{\xi}_{nlm}(\vec{r})$ form an orthogonal basis for a Hilbert space with the inner product
\begin{equation}
\langle\vec{\eta},\vec{\beta}\rangle = \int_{\mathcal{V}} \rho \vec{\eta}\cdot \vec{\beta}\, d^3\vec{r},
\label{inner}
\end{equation}
where $\vec{\eta}$ and $\vec{\beta}$ are arbitrary vector functions defined over the volume of the crust $\mathcal{V}$. Therefore, an arbitrary solenoidal displacement field of the crust $\vec{\xi}(\vec{r},t)$ may be decomposed as
\begin{equation}
\label{xidef}
\vec{\xi}(\vec{r},t) = \sum_{n,l,m} a_{nlm}(t)\,\vec{\xi}_{nlm}(\vec{r}),
\end{equation} 
where
\begin{equation}
a_{nlm}(t) = \frac{\langle\vec{\xi}(\vec{r},t),\vec{\xi}_{nlm}\rangle}{\langle\vec{\xi}_{nlm} ,\vec{\xi}_{nlm}\rangle}.
\label{coefficient}
\end{equation}
Effectively, the spectral method replaces the crust with many oscillators.
Equation \eqref{xieqn} describes free oscillations, with no external forces, and is reduced to $\ddot{a}_{nlm}(t) + \omega_{nlm}^2 a_{nlm}(t)=0$. In the presence of
magnetic coupling to the magnetosphere/core, external forces $\vec{f}_\text{mag}$ and $\vec{f}_\text{core}$ appear at the upper/lower boundaries of the crust,
\begin{equation}
  \vec{f}_\text{ext}= \vec{f}_\text{mag} + \vec{f}_\text{core}.
\end{equation}
Then each oscillator is driven by the projection of the external force on the eigenmode,
\begin{equation}
    \ddot{a}_{nlm}(t) + \omega_{nlm}^2 a_{nlm}(t) = \frac{\langle \vec{f}_{\text{ext}}(\vec{r},t),\vec{\xi}_{nlm}\rangle}{\langle\vec{\xi}_{nlm} ,\vec{\xi}_{nlm}\rangle}.
\label{a_eqn}
\end{equation}
The initial conditions $a_{nlm}(t=0)$ are determined by the initial displacement $\vec{\xi}_0$ and Equation~\eqref{coefficient}. We then evolve the spectral coefficients $a_{nlm}$, our effective dynamical variables, using Equation \eqref{a_eqn}. 
\medskip

\subsection{Basis Functions} \label{basis_functions}

For the class of solenoidal displacements we are considering, and the above operators, the natural choice of basis functions is
\begin{equation}
\vec{\xi}_{nlm} = \ff_{nl}(r)\,\vec{r}\times\nabla Y_{lm},
\label{basis}
\end{equation}
where $\ff_{nl}$ contains the radial part of the eigenfunction, and $\vec{r}\times\nabla Y_{lm}$ is the third vector spherical harmonic. Substitution of Equation \eqref{basis} into Equation \eqref{xieqn} results in the following Sturm-Liouville problem:
\begin{equation}
\begin{aligned}
-\omega_{nl}^2 \rho \ff_{nl} =  &\frac{d\tilde{\mu} }{dr}\bigg(\frac{d\ff_{nl}}{dr} - \frac{\ff_{nl}}{r} \bigg) + \frac{\tilde{\mu}}{r^2}\frac{d}{dr}\bigg(r^2\frac{d\ff_{nl}}{dr}\bigg)\\
& -[ l(l+1)\mu + 2\mu_B]\frac{\ff_{nl}}{r^2},
\end{aligned}
\label{f_eqn}
\end{equation}
The radial eigenfunctions $\ff_{nl}(r)$ and eigenvalues $\omega_{nlm}=\omega_{nl}$ do not depend on the azimuthal mode number $m$ due to the spherical symmetry of $\tilde{\mu}$. 
Note that in the limit $\mu_B\rightarrow 0$ Equation~\eqref{f_eqn} is the same as Equation~(23) in \cite{mcdermott_nonradial_1988}. 

We use a high-order Sturm-Liouville solver to numerically find the eigenfunctions and eigenvalues of Equation \eqref{f_eqn}. The details are given in Appendix \ref{elastic_modes}. 
\medskip

\subsection{Coupling to the Core}\label{crust_core}
The magnetic field is frozen in the crust and the liquid core, and so crustal oscillations deform the magnetic field lines and launch Alfv\'en waves into the core. The feedback of these waves on the crust dynamics is incorporated in our simulations as follows.\\

For simplicity, we approximate the background magnetic field $\vec{B}$ as purely radial so that $B=B_r$. Since the core is effectively an infinite reservoir on the quake timescale (Section~\ref{core}), there are only inward propagating waves with the displacement of the form $\vec{\xi}(t+r/v_A)$, where $v_A$ is the Alfv\'en speed in the core. 
The magnetic field of the emitted waves is related to the displacement $\vec{\xi}$ by the flux-freezing condition, 
\beq
    \delta\vec{B}_< = \nabla\times (\vec{\xi}\times\vec{B})
    =\frac{1}{r}\,\partial_r(B_r r\,\vec{\xi})
    \approx B_r\partial_r\vec{\xi}=\frac{B_r}{v_A}\,\dot{\vec{\xi}}.
\eeq
Here subscript ``$<$'' stands for the core region immediately below the crust, and $\dot{\vec{\xi}}$ is the time derivative of the displacement at the interface.

The presence of $\delta \vec{B}_<$ implies that the core applies Maxwell stress to the bottom of the crust. The extracted momentum flux is
\begin{equation}
\sigma_{rh} = - \frac{B_r\, \delta B_{h,<}}{4\pi},
\end{equation}
where $h=\theta, \phi$ labels the horizontal component. Since the crustal modes are calculated with the stress-free boundary condition $\delta\vec{B}=0$, the external stress must be included as a driving term in the oscillation Equation~(\ref{a_eqn}). The external force appearing in this equation is applied to the bottom layer of the crust of some thickness $\Delta r$ and density $\rho_>$ (just above the interface), so that $f_{\rm ext}\rho_> \approx \sigma_{rh}/\Delta r$. Approximating the layer as infinitesimally thin, the external force at the crust-core interface becomes
\begin{equation}
  \vec{f}_\text{core} = - \frac{B_r\,\delta\vec{B}_<}{4\pi\rho_>}\,\delta(r-r_c).
\end{equation}
Substituting the core Afv\'en speed $v_A=B_r/(4\pi\rho_<)^{1/2}$, we obtain
\begin{equation}
 \vec{f}_\text{core} = -\frac{\rho_<}{\rho_>}\,v_A\,\dot{\vec{\xi}}\,\delta(r-r_c),
 \label{f_core}
 \end{equation}
where $\rho_<$ is the mass density of the core infinitesimally below the crust-core interface. One can see that coupling to the core is equivalent to adding a damping force $\propto \dot{\vec{\xi}}$. 

The projection of $\vec{f}_\text{core}$ onto each basis function is computed once at the beginning of the simulation and stored in an array (see Appendix \ref{numerical_method_crust}).
\medskip

\subsection{Coupling to the Magnetosphere} \label{dynamics_magnetosphere}

In this work, we model the pulsar magnetosphere as dipole, and treat the magnetospheric waves as linear perturbations, using the framework of force-free electrodynamics. In force-free electrodynamics the inertia of the plasma is negligible compared to the inertia of the magnetic field, and the equation of motion is replaced by the condition
\begin{equation}
\rho_\text{e}\vec{E} + \frac{\vec{j}\times\vec{B}}{c} = 0.
\label{force}
\end{equation}
It implies $\vec{E}\cdot\vec{B}=0$ and $\vec{E}\cdot\vec{j}=0$, so there is no dissipation. This approximation is valid if there is enough plasma to sustain electric currents excited in the perturbed magnetosphere. For linear perturbations about a stationary background state with $\vec{E}=0$ (in the corotating frame) and $\nabla\times\vec{B}=0$ the force-free condition becomes $\delta\vec{j}\times\vec{B}=0$. Substitution of $\delta\vec{j}$ from Equation~(\ref{eq:dj}) then gives
\begin{equation}
  \rho_B\,
  \ddot{\vec{\xi}}_\perp
  = \frac{1}{4\pi}[\nabla\times\nabla\times(\vec{\xi}\times\vec{B})]\times\vec{B}.
\label{alfven}
\end{equation}
Note that only the perpendicular displacement $\vec{\xi}=\vec{\xi}_\perp$ enters the force-free wave equation. 

The wave equation gives the dispersion relation for eigen modes $\vec{\xi}\propto \exp(-i\omega t+\vec{k}\cdot\vec{r})$,
\beq
    \frac{\omega^2}{c^2}\,\vec{\xi}=k_\parallel^2\,\vec{\xi}
  +\vec{k}_\perp (\vec{k}\cdot\vec{\xi}),
\eeq 
where $k_\parallel$ and $\vec{k}_\perp$ are the components of $\vec{k}$ parallel and perpendicular to $\vec{B}$, respectively.
The eigen modes include shear Alfv\'en waves ($\vec{k}\cdot{\vec{\xi}}=0$) with dispersion relation $\omega=k_\parallel c$, and compressive (called ``fast'') modes. The perturbations are generated by the shear motions of the crust at the footprints of the magnetospheric field lines, and these motions should launch Alfv\'en waves. Their conversion to fast modes in the magnetosphere is a second-order effect, which is negligible as long as $\delta B/B \ll 1$.

 The group speed of Alfv\'en waves is parallel to $\vec{B}$, so they are ducted along the magnetic field lines. For the linear dynamics of Equation \eqref{alfven}, each poloidal field line behaves like an independent string, with no coupling to other field lines. Then effectively we need to solve a 1D wave equation along each poloidal field line.

In axisymmetry, $\partial/\partial\phi=0$, the Alfv\'en waves have the displacement in the $\phi$-direction, $\vec{\xi}=\xi_\phi\,\hat{\vec{\phi}}$. It is convenient to work in the so-called magnetic flux coordinates $(\psi,\chi,\phi)$. The coordinate $\psi$ represents surfaces of constant poloidal flux (for a dipole magnetosphere it is given by Equation~(\ref{eq:psi})), and $\chi$ is the length along poloidal field lines in the $\phi=\text{const}$ plane \citep{goedbloed_advanced_2010}. Equation \eqref{alfven} can be written in the flux coordinates as
\begin{equation}
\frac{\partial^2 \xi_\phi(\psi,\chi)}{\partial t^2} = 
\frac{c^2}{r_\perp B}\,
\frac{\partial}{\partial \chi}\left[ r_\perp^2 B \frac{\partial}{\partial \chi}\left(\frac{\xi_\phi(\psi,\chi)}{r_\perp}\right)\right],
\label{alfven_flux}
\end{equation}
where $r_\perp=r\sin\theta$ is the cylindrical radius. Each flux surface in the magnetosphere is  effectively a 1D string (with mass density and tension both proportional to $Br_\perp$) supporting shear wave propagation with speed $c$. 

Between the solid crust and the force-free magnetosphere there is the liquid ocean. The ocean dynamics can be calculated by extending the magnetosphere model so that each 1D string includes a heavy part at the footprint where the string mass density is increased and the shear wave is decelerated below $c$ as $v_A/c=(\rho/\rho_B+1)^{-1/2}$. The technical motivation for treating the ocean motions as part of the magnetospheric dynamics is that it is liquid and hence ``force-free'' --- it does not sustain any shear forces. Note however that the ocean depth is small compared with the crust thickness, and at wave frequencies of  interest, it moves together with the crust at the footprints of the magnetospheric field lines. Effectively, the magnetosphere is attached to the solid crust, and in the numerical models presented in Section~\ref{results} the presence of the ocean will be neglected. We also performed more detailed simulations with ocean dynamics included, which support this approximation for Vela.

Solving the magnetospheric field-line dynamics requires two boundary conditions. For closed field lines, the boundary conditions are applied at the two footpoints where the field line intersects with the surface of the neutron star. The field line is attached to the star and its footprint displacement equals the instantaneous displacement of the uppermost layer of the crust, $\vec{\xi}(t,r_\star)$, which is determined by Equation \eqref{a_eqn}. 

For open field lines, only one end is attached to the star, giving one boundary condition $\vec{\xi}(r_\star)$. The other end is at the outer boundary of the computational domain. At this end, we apply the condition of free escape, which means that there are only outgoing Alfv\'en waves. Outgoing waves are functions of $t-\chi/c$ and satisfy the condition
\begin{equation}
\label{eq:open}
  \frac{\partial\vec{\xi}}{\partial\chi}\bigg|_{\chi_{\text{end}}} = -\frac{1}{c}\frac{\partial\vec{\xi}}{\partial t}\bigg|_{\chi_{\text{end}}},
\end{equation}

In our simulations, the magnetosphere is sampled with 275 closed and 50 open flux surfaces. The outer boundary of the open field lines is set at $r_\text{max}=10^7$ cm, and the last closed field line extends to $R_\text{LC}=4.2\times 10^8$ cm --- the light cylinder radius of Vela. We follow the dynamics of each field line by solving the string Equation~\eqref{alfven_flux} with the boundary condition $\xi(r_\star)$ at the footprints and Equation~(\ref{eq:open}) at the outer boundary. The magnetospheric dynamics is coupled to the crustal oscillations at $r_\star$, so the crust and the magnetosphere evolve together as a coupled system. The coupled differential Equations \eqref{a_eqn} and \eqref{alfven_flux} are integrated numerically using the fourth-order Runge-Kutta scheme, as described in Appendices \ref{numerical_method_crust} and \ref{numerical_method_magnetosphere}.

The feedback of the emitted magnetospheric waves on the crust oscillations is implemented similarly to the crust-core interaction described in Section~\ref{crust_core}. In the axisymmetric model, both the displacement and the perturbed magnetic field are in the $\phi$-direction. Let $\delta B=B_{\phi,>}$ be the perturbed field immediately above the stellar surface. The magnetospheric stress $B_rB_{\phi,>}/4\pi$ is communicated directly to the solid crust at the bottom of the ocean, where density $\rho=\rho_{\rm crys}$. To extract the required momentum flux $\sigma_{r\phi}=-B_rB_{\phi,>}/4\pi$ from the crust, we apply force $f_{\rm mag}=-(\sigma_{r\phi}/\rho_{\rm cryst} \Delta r)$ to the upper layer of the solid material with a small thickness $\Delta r$,
\begin{equation}
   \vec{f}_\text{mag} \approx \frac{B_r\,\delta\vec{B}_>}{4\pi\rho_{\rm crys}}\,\delta(r-r_{\rm crys}).
\end{equation}
The magnetospheric perturbation $B_\phi$ is related to the displacement $\xi_\phi(\psi,\chi)$ by the flux-freezing condition,
\begin{equation}
   \delta B_\phi 
   = B r_\perp\frac{\partial}{\partial\chi}\left( \frac{\xi_\phi}{r_\perp}\right).
\end{equation}

This allows one to express $\vec{f}_{\rm mag}$ in the form
\begin{equation}
   f_\text{mag}^\phi = \frac{\rho(r_>)}{\rho_{\rm crys}} v_A^2(r_>) \cos\alpha\, \,r_\perp \frac{\partial}{\partial\chi}\left( \frac{\xi_\phi}{r_\perp}\right)\bigg|_{r_>} \,\delta(r-r_\star),
\label{mag_force_phi}
\end{equation}
where $\alpha$ is the angle between the magnetic flux surface and the radial direction.  In the model where the magnetosphere is directly attached to the solid crust (neglecting the thin ocean), $v_A(r_>)=c$ and $\rho(r_>)=\rho_B$. This approximation is used in the simulations presented below. A more detailed model of magnetospheric waves with the ocean at the footprints would have $v_A(r_>)\approx B/(4\pi \rho_{\rm crys})^{1/2}\approx 10^{-2}c$ and $\rho(r_>)=\rho_{\rm crys}$. It would explicitly follow the wave acceleration to $c$ as it crosses the ocean.
\medskip


\section{Sample models}\label{results}

\begin{table*}
\centering
\caption{Sample models.}
\begin{tabular*}{\textwidth}{c @{\extracolsep{\fill}} ccccccc}
\hline
Model & Quake Location & Core  & Core $v_A$ & $ \rho_< / \rho_> $ & $\epsilon_0$ & $A_0$ & $E_Q$ \\
\hline
A1 & Polar cap      & Decoupled       & ---                       & --- & $4.4\times10^{-4}$ & $3\times10^{11}$ cm$^2$ & $10^{38}$ erg \\
A2 & Polar cap      & Superconducting & $5\times10^6$ cm s$^{-1}$ & 0.1 & $4.4\times10^{-4}$ & $3\times10^{11}$ cm$^2$ & $10^{38}$ erg \\
B1 & $\theta=\pi/4$ & Decoupled       & ---                       & --- & $1.3\times10^{-4}$ & $1\times10^{12}$ cm$^2$ & $10^{38}$ erg \\
B2 & $\theta=\pi/4$ & Superconducting & $5\times10^6$ cm s$^{-1}$ & 0.1 & $1.3\times10^{-4}$ & $1\times10^{12}$ cm$^2$ & $10^{38}$ erg \\
\hline
\end{tabular*}
\label{Sample_models}
\end{table*}

\begin{figure}%
\centering
\includegraphics[width=0.42\textwidth]{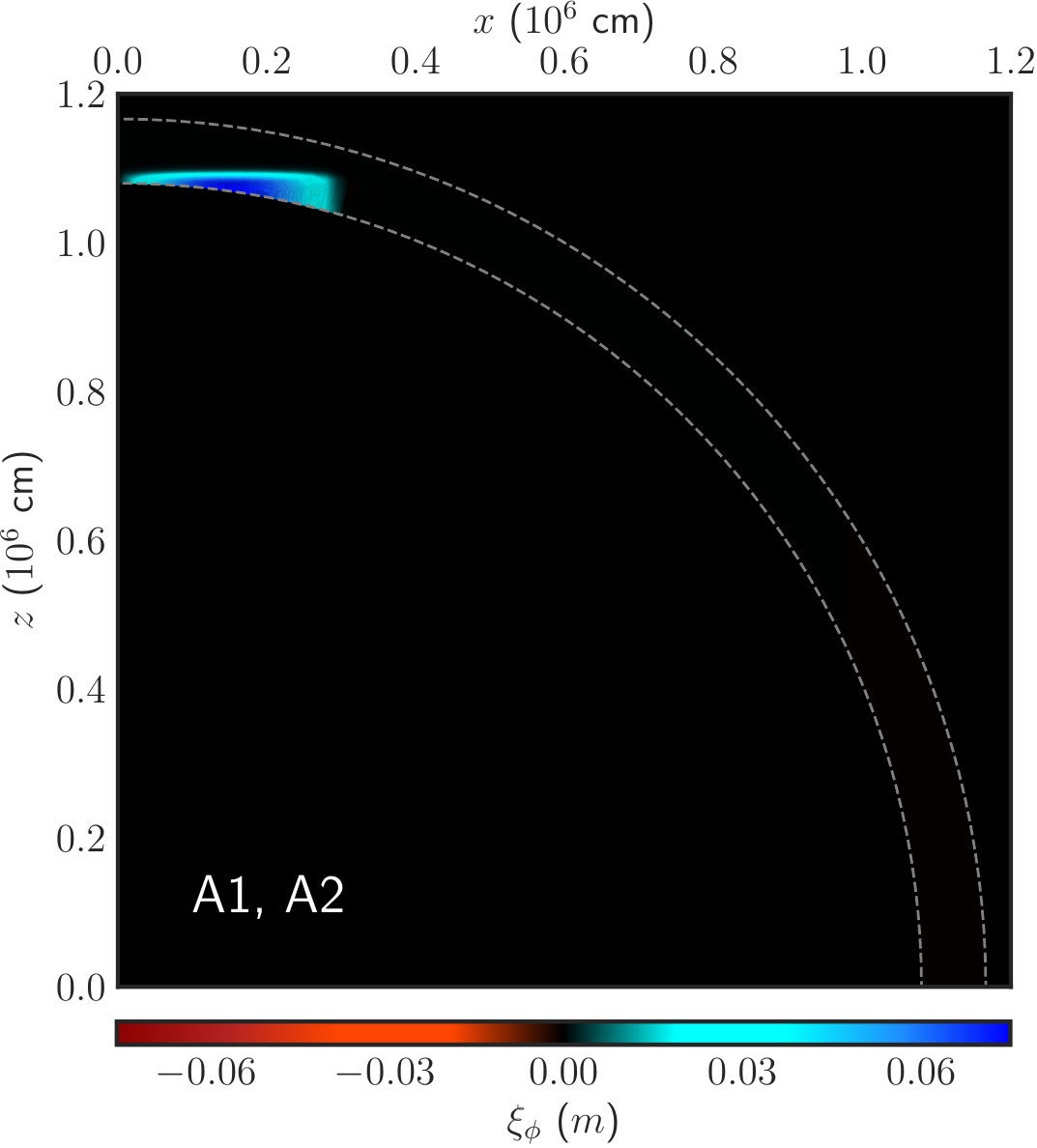} 
\includegraphics[width=0.42\textwidth]{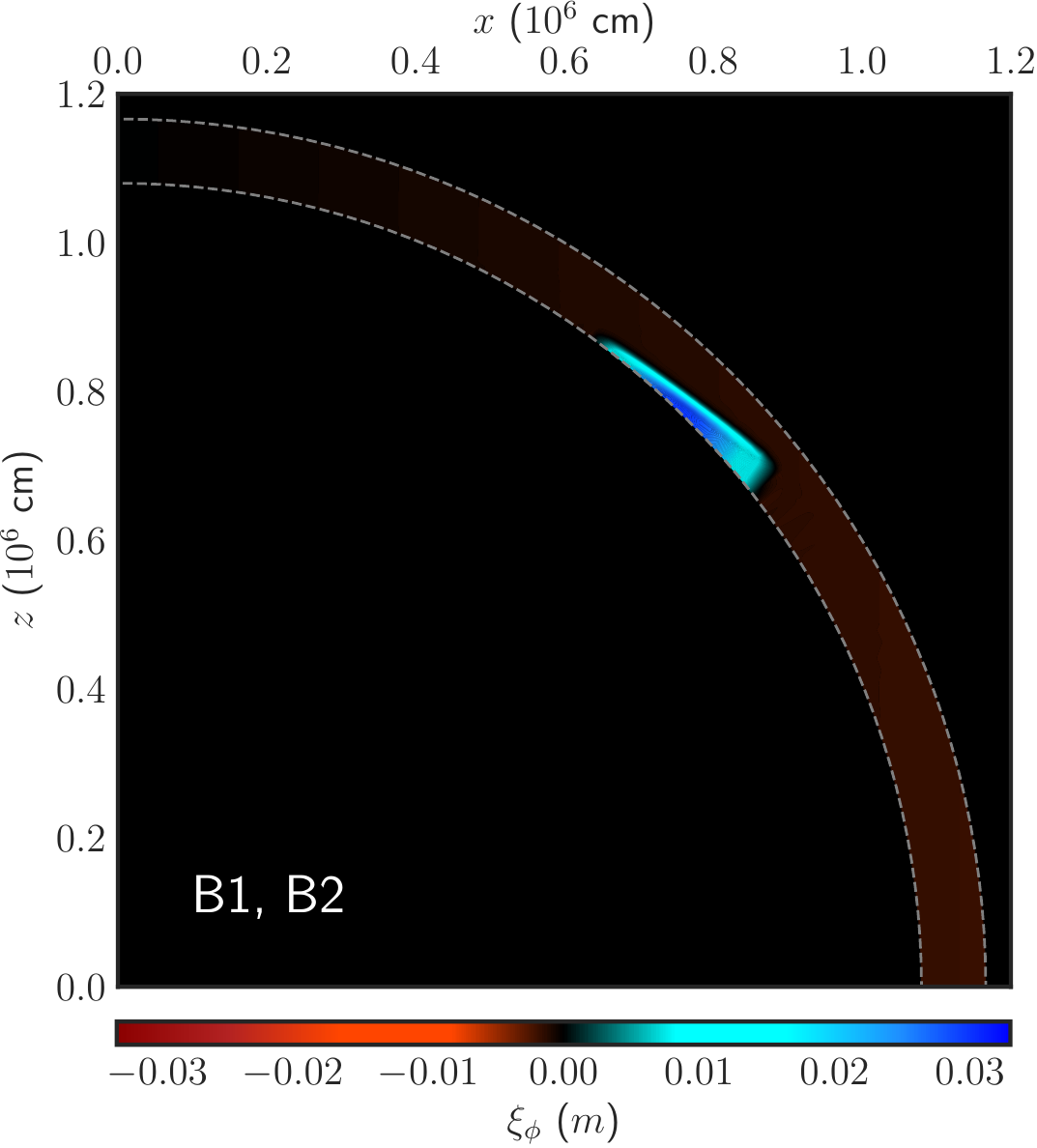}
\caption{Top: initial conditions used for models A1 and A2. Bottom: initial conditions used for models B1 and B2. Color shows the amplitude of the azimuthal displacement $\xi_\phi$. The amplitude is scaled so that each initial condition has the initial energy $E=10^{38}$ erg. The gray dashed lines show the boundaries of the crust.}
    \label{ICs}
\end{figure}

We have calculated four sample models: A1, B1, and A2, B2. Their parameters are given in Table \ref{Sample_models}, and the initial displacement of the disturbed crust is shown Figure \ref{ICs}. In all the models, the quake has energy $E_Q=10^{38}\,$erg.

Models A1 and B1 have no crust-core coupling, representing a pulsar with a magnetic field confined to the crust and not penetrating the core. Models A2 and B2 have strong crust-core coupling; they assume a superconducting core, and the poloidal component of the magnetic field at the crust-core interface $B\approx 3.4\times 10^{12}\,$G, similar
to the  measured surface dipole field of Vela. 

\begin{figure*}
\centering
\includegraphics[height=0.309\textwidth]{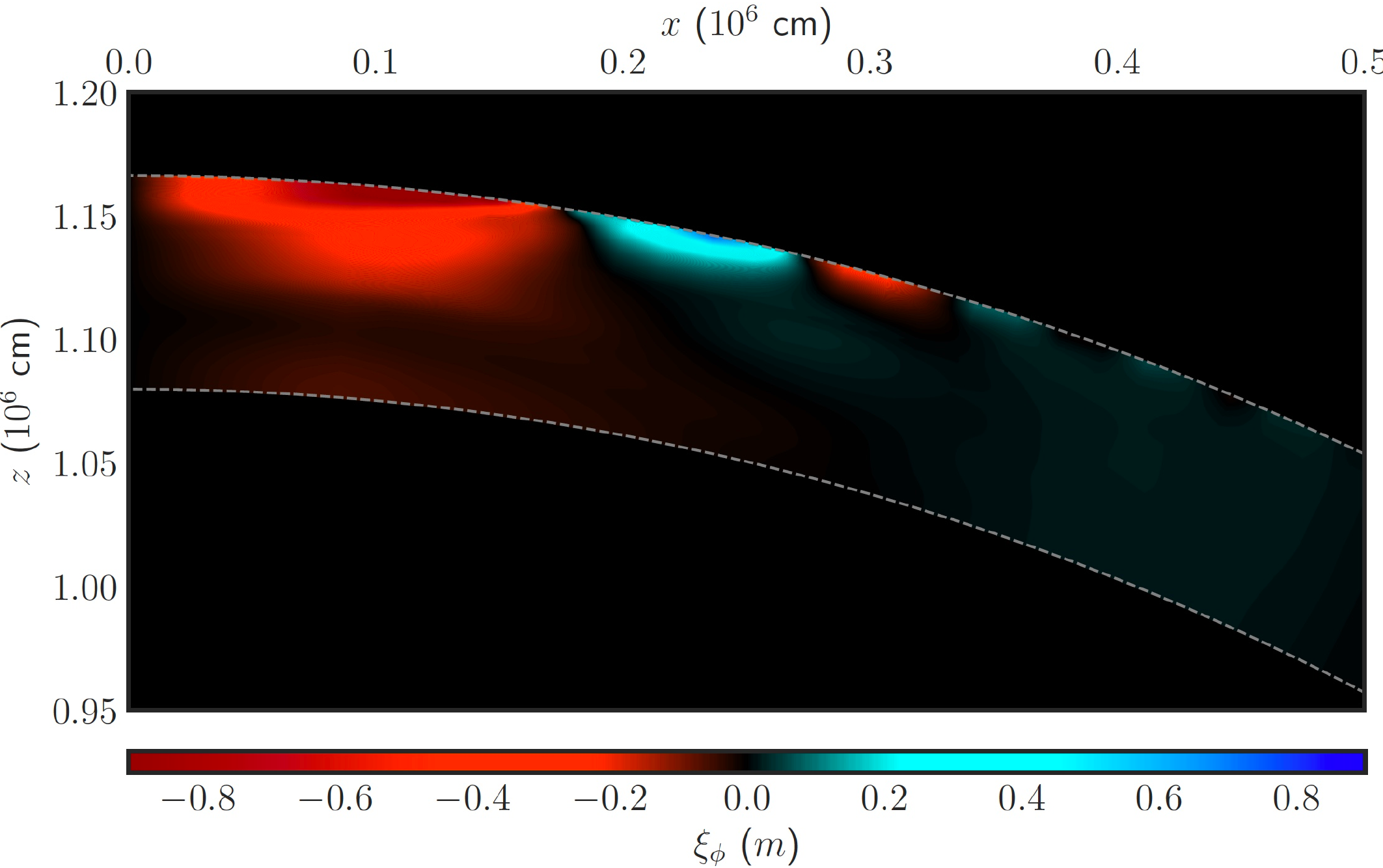} 
\includegraphics[height=0.31\textwidth]{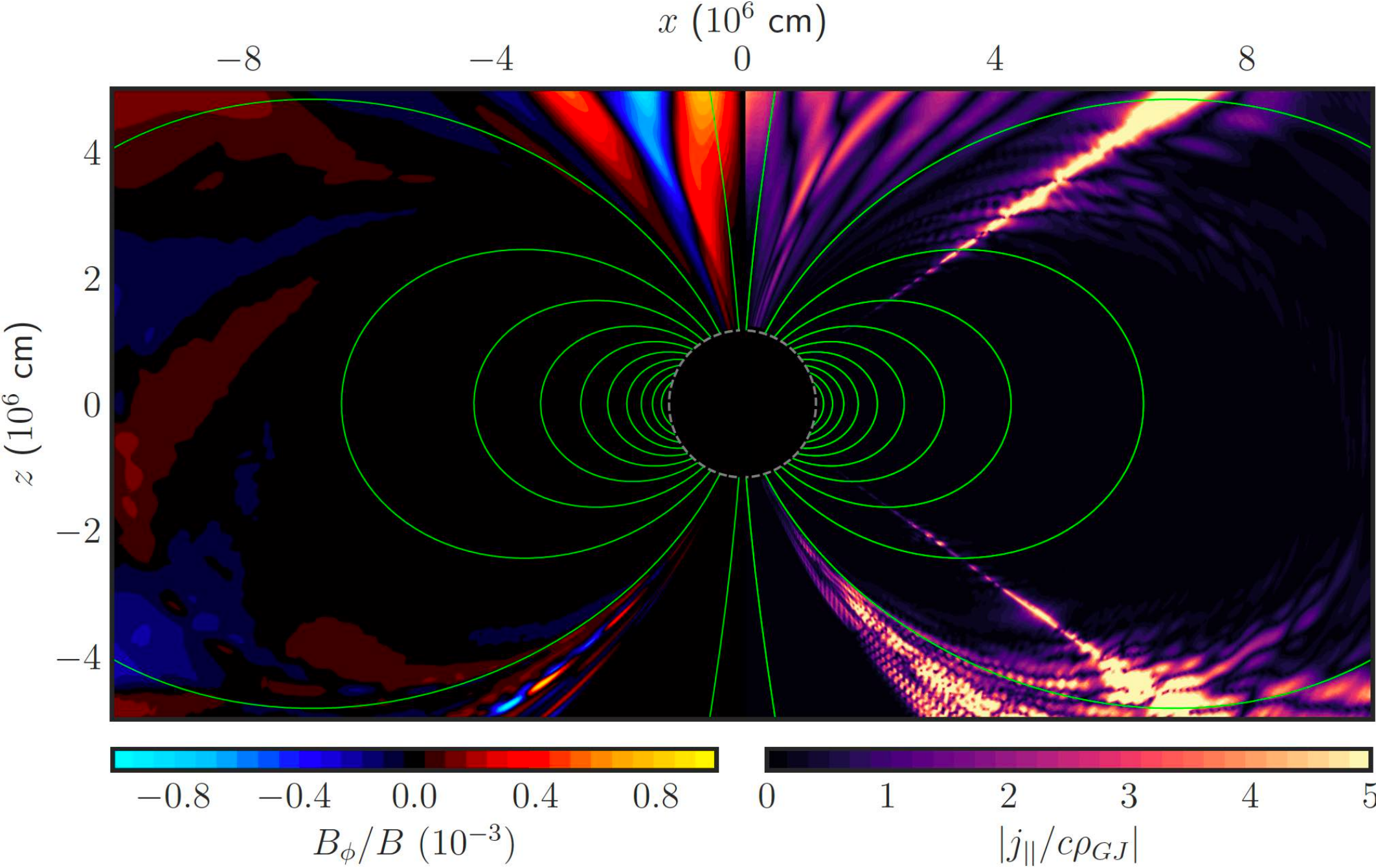} 
\hspace*{-.11in}
\includegraphics[width=1.01\textwidth]{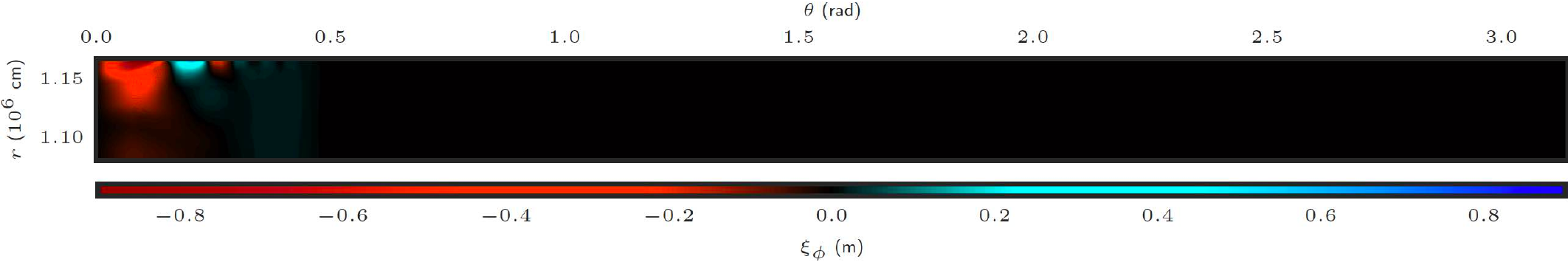}
    \caption{Model A1 at $t=2$ ms. Top left: displacement $\xi_\phi$ of the crust near the epicenter of the quake. The dashed lines show the boundaries of the crust.
Top right: toroidal perturbation of the magnetic field $B_\phi / B$ (left), and the ratio 
$|j_\parallel / c \rho_{GJ}|$ (right). The green 
curves
show the poloidal magnetic field. The 
two
field lines closest to the axis of symmetry 
are 
the boundary of the open field-line bundle.
The gray dashed 
circle
is the surface of the neutron star. Bottom: displacement $\xi_\phi(r,\theta)$ in the entire crust, plotted on the $r$-$\theta$ plane.
}
    \label{A1}
\end{figure*}

\begin{figure*}
\centering
\includegraphics[height=0.309\textwidth]{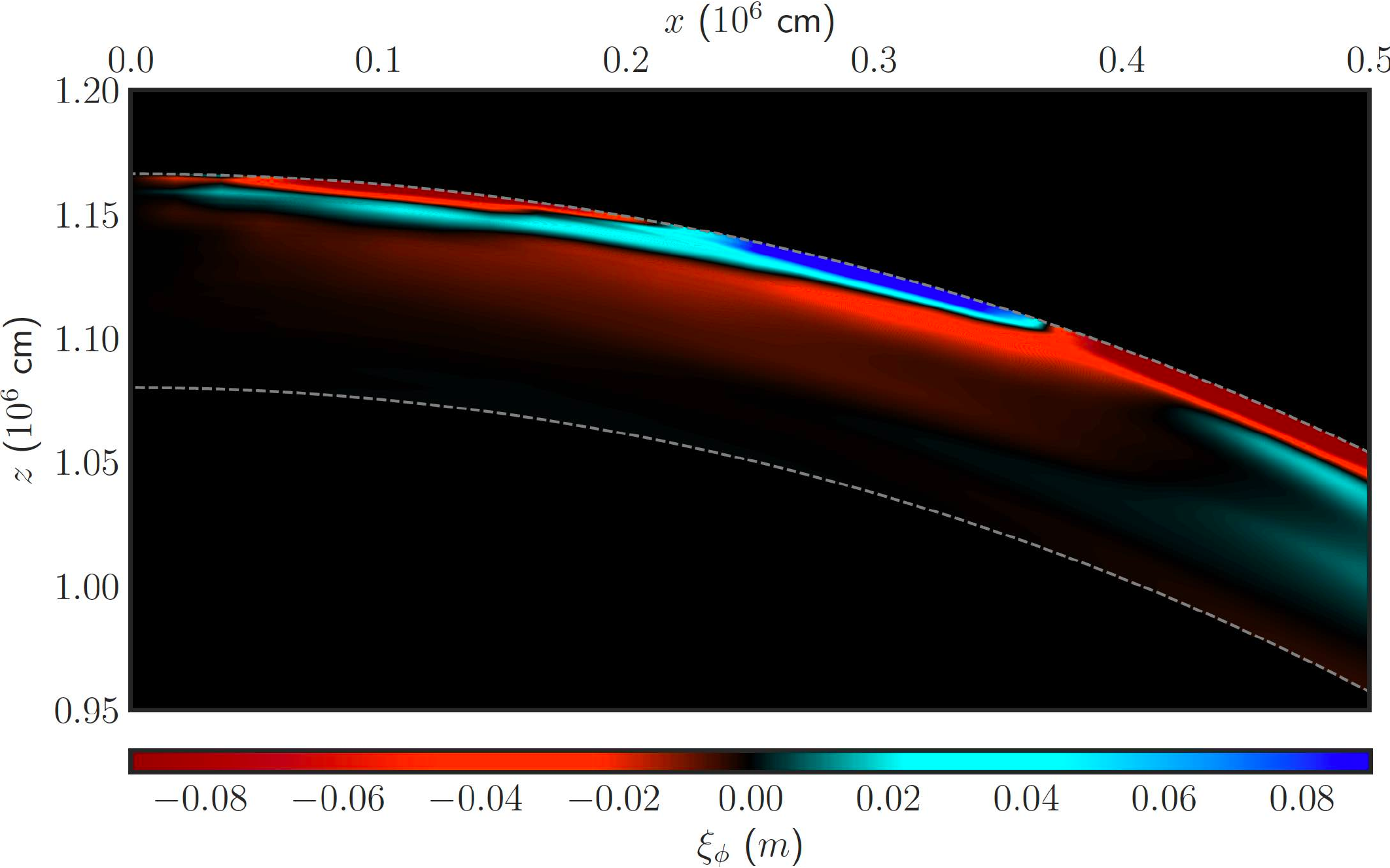} 
\includegraphics[height=0.31\textwidth]{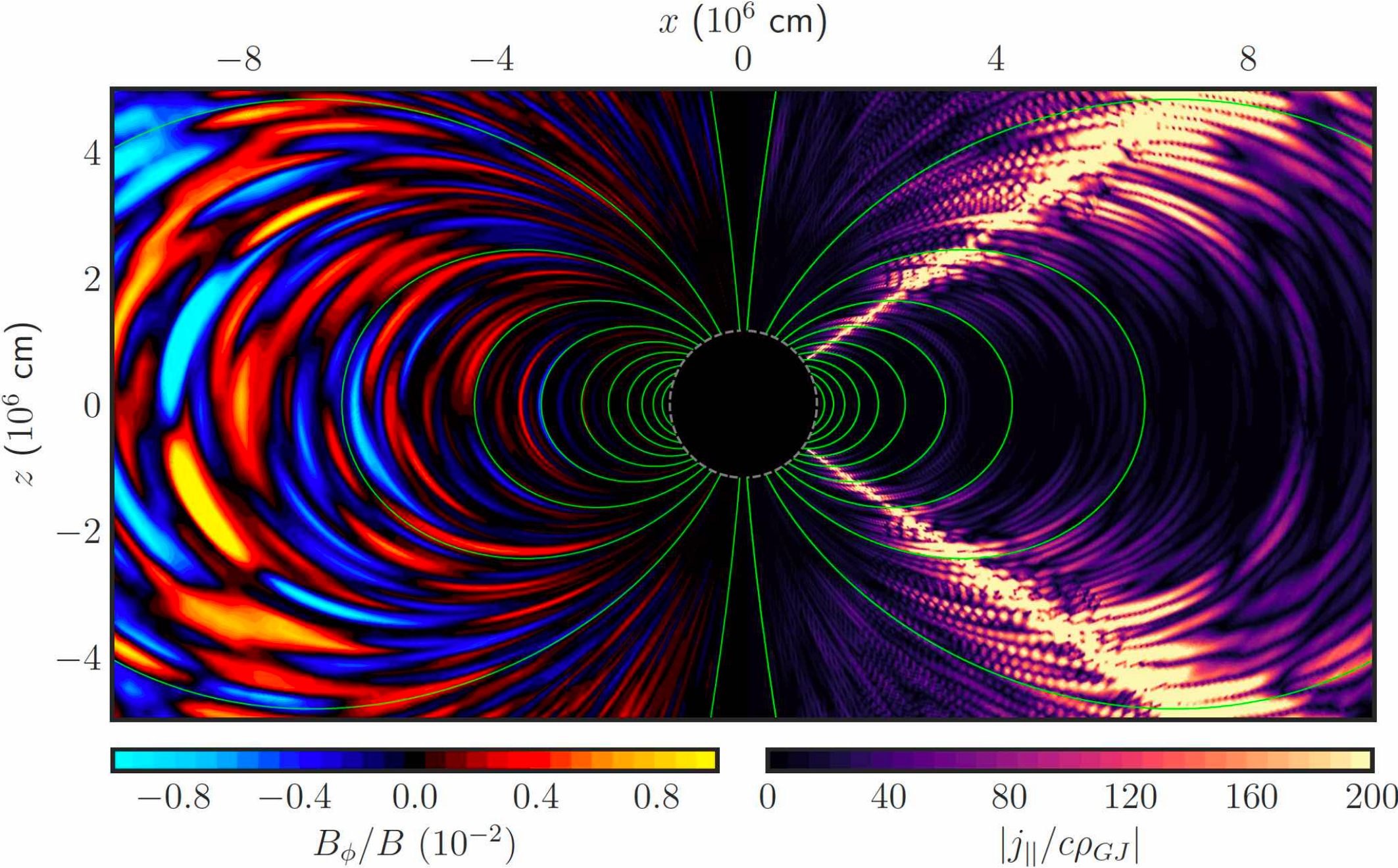} 
\hspace*{-.18in}
\includegraphics[width=1.01\textwidth]{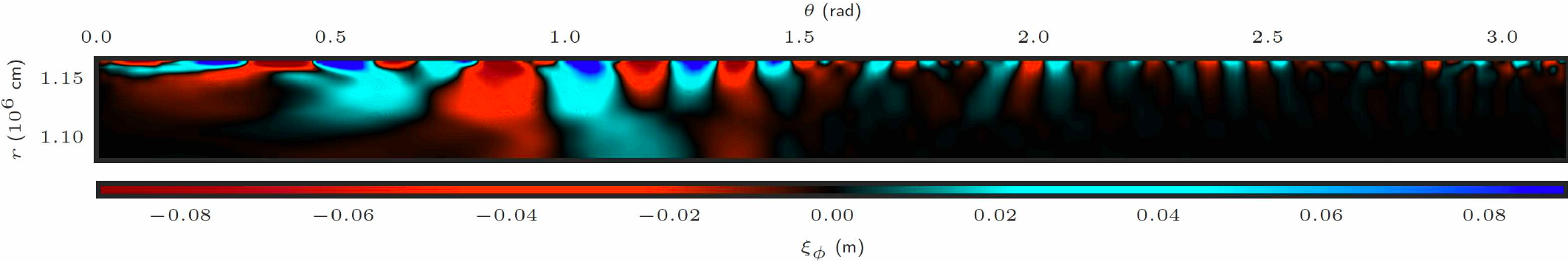} 
    \caption{Same as Figure~\ref{A1} but at time $t=50\,$ms.
    }
    \label{A2}
\end{figure*}

\begin{figure*}
\centering
\includegraphics[width=.41\textwidth]{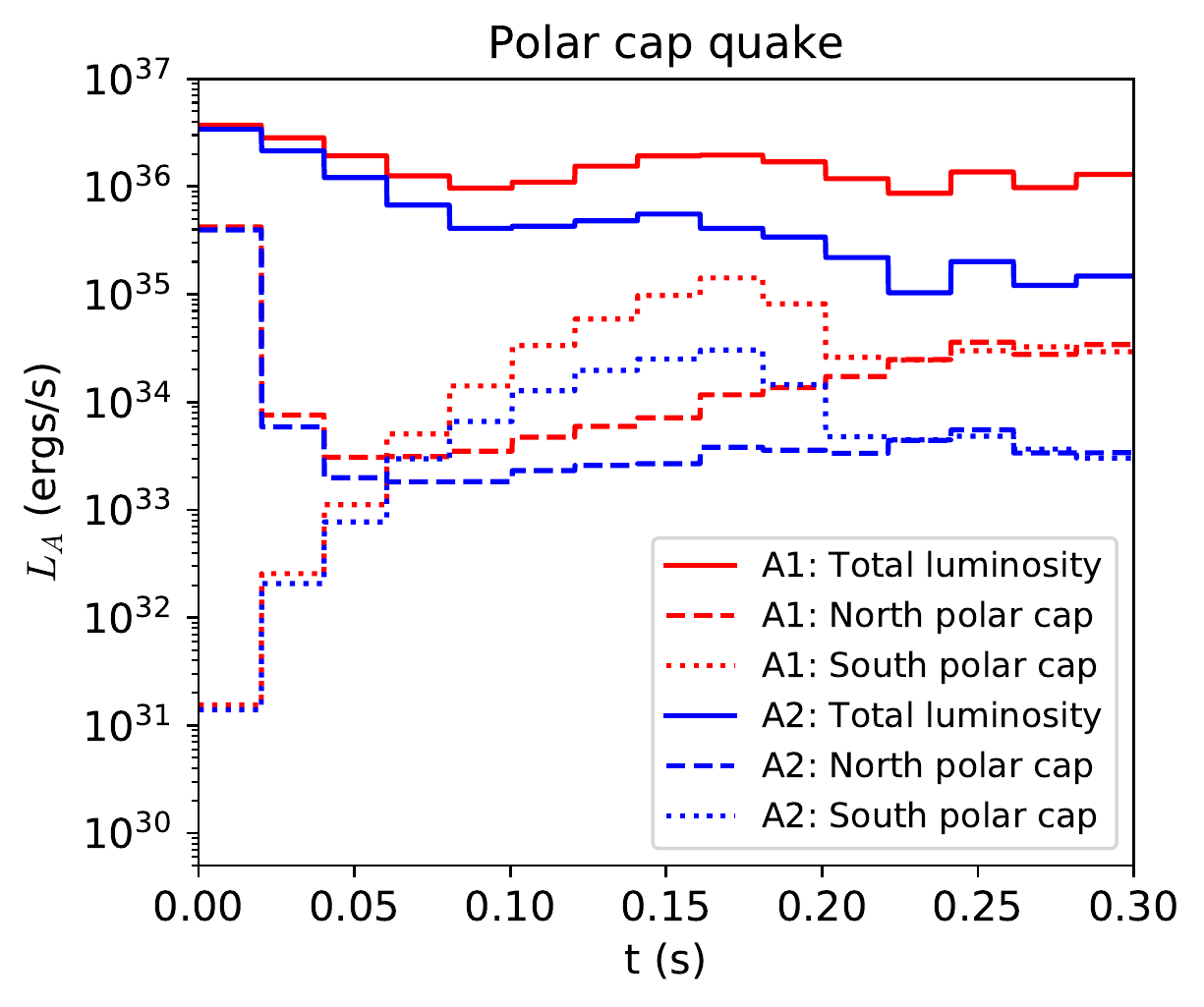}
\includegraphics[width=.41\textwidth]{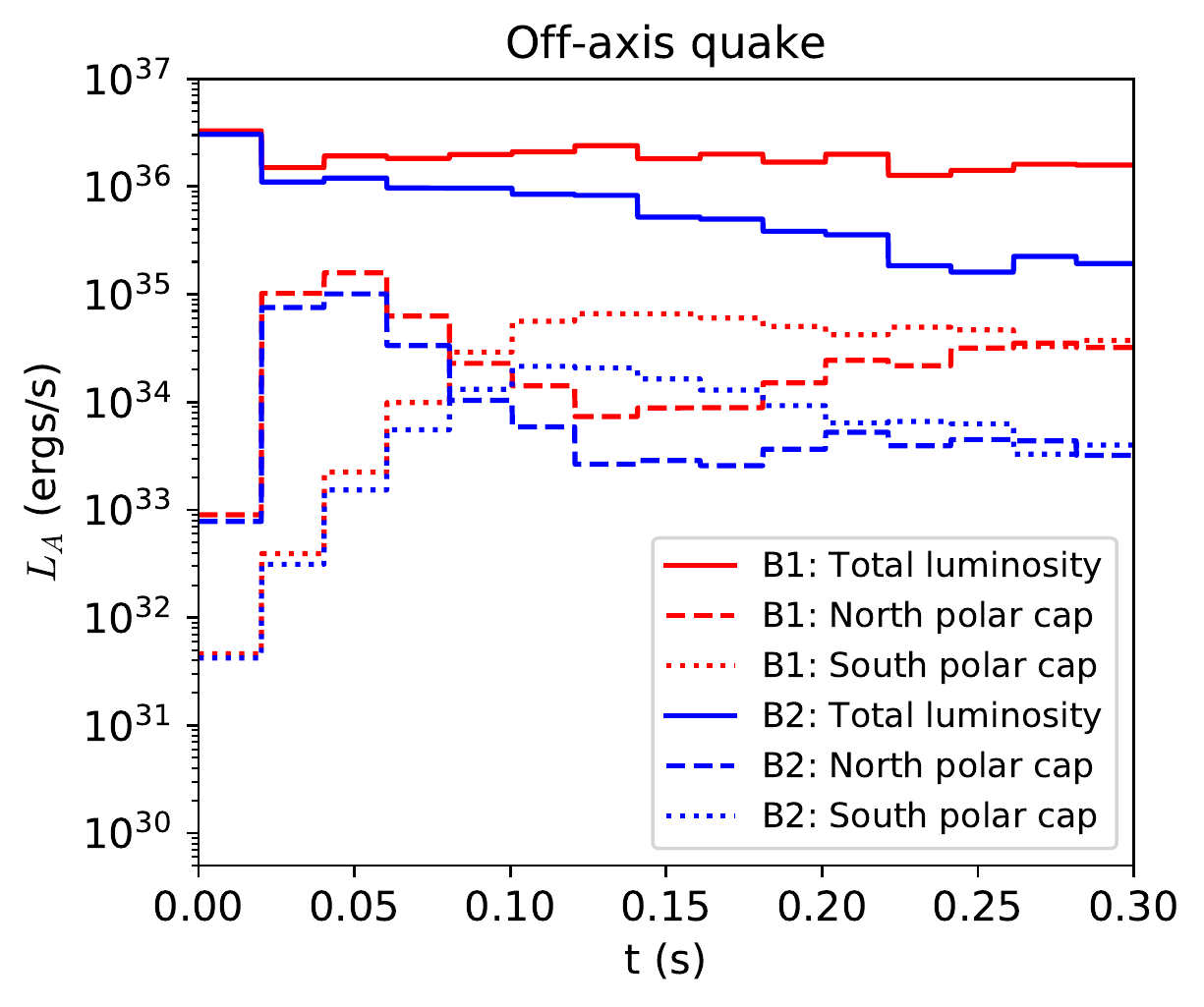}
\caption{Luminosity of Alfv\'en waves emitted into the magnetosphere, $L_{\rm A}$. The luminosity has been averaged into $20$ ms bins to remove the noise from fast oscillations. Left: models A1 and A2 (initial quake under the polar cap). Right: models B1 and B2 (initial quake at $\theta\sim \pi/4$). Red is used for models with no crust-core coupling (A1 and B1), and blue for models with strong crust-core coupling (A2 and B2). For each model, we show $L_{\rm A}$ from the entire stellar surface (solid curve), and the contributions from the north (dashed) and south (dotted) polar caps.} 
\label{lum}
\end{figure*}

The dynamical picture of quake development is quite similar in all four models. As an example, the snapshots of model~A1 are shown in Figures \ref{A1} and \ref{A2}. At the beginning, we observe shear waves propagating toward the surface and launching Alfv\'en waves into the magnetosphere directly above the quake region (which is at the north polar cap in model A1). Due to the large impedance mismatch at both
the crust-core and the crust-magnetosphere interfaces, most of the quake energy remains trapped inside the crust, and the waves bounce many times between the two interfaces. Some waves are launched in the $\hat{\theta}$-direction with a large surface amplitude and cross the circumference of the crust in a time $\pi r_\star/\tilde{v}_s\sim 30\,$ms. These surface waves are the so-called ``whispering gallery modes" \citep{rayleigh_theory_1894}. However, most of the shear wave energy remains concentrated at the north pole for a longer time, and gradually spreads toward the south pole after many small angle reflections at the interfaces. As the centroid of the shear wave energy passes the magnetic equator the luminosity of Alfv\'en waves into the magnetosphere, $L_{\rm A}$, drops because $B_r$ is small. After $\sim 200$ ms the wave energy has spread throughout the entire crust, and the same luminosity of Alfv\'en waves is measured from the north and south poles. The evolution of $L_{\rm A}$ is shown in Figure \ref{lum}.

\begin{figure*}
\centering
\includegraphics[width=.40\textwidth]{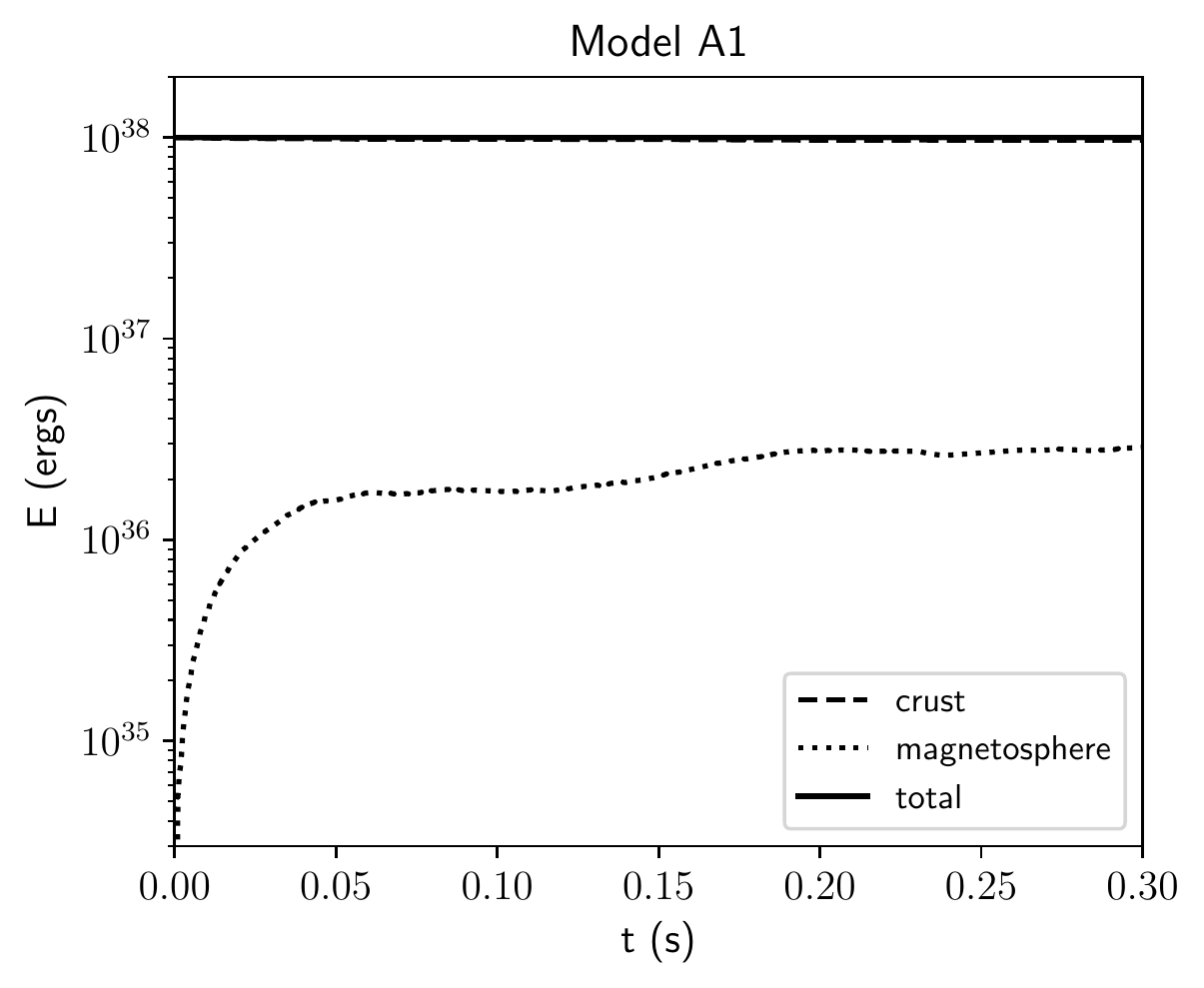}
\includegraphics[width=.40\textwidth]{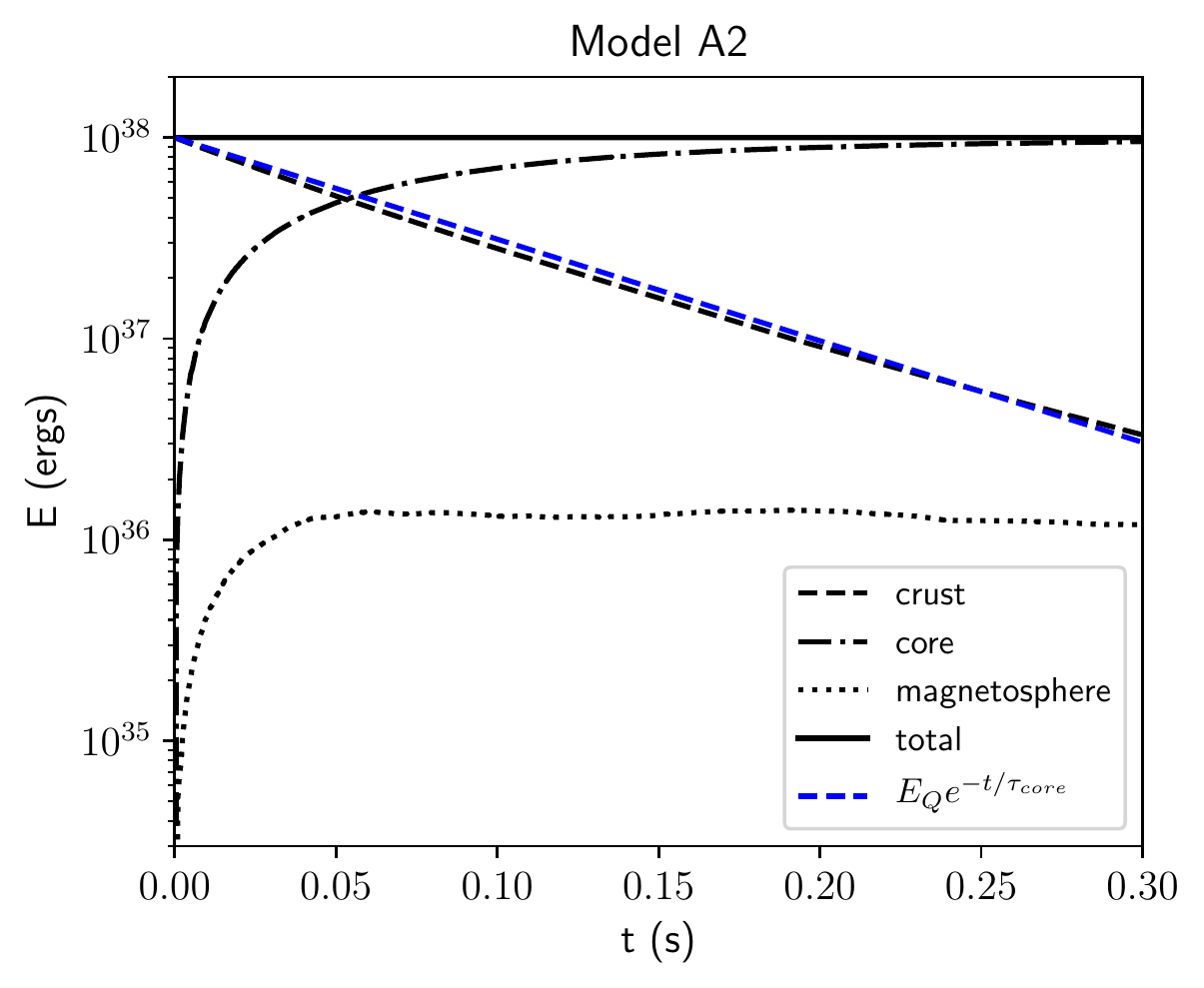}
\includegraphics[width=.40\textwidth]{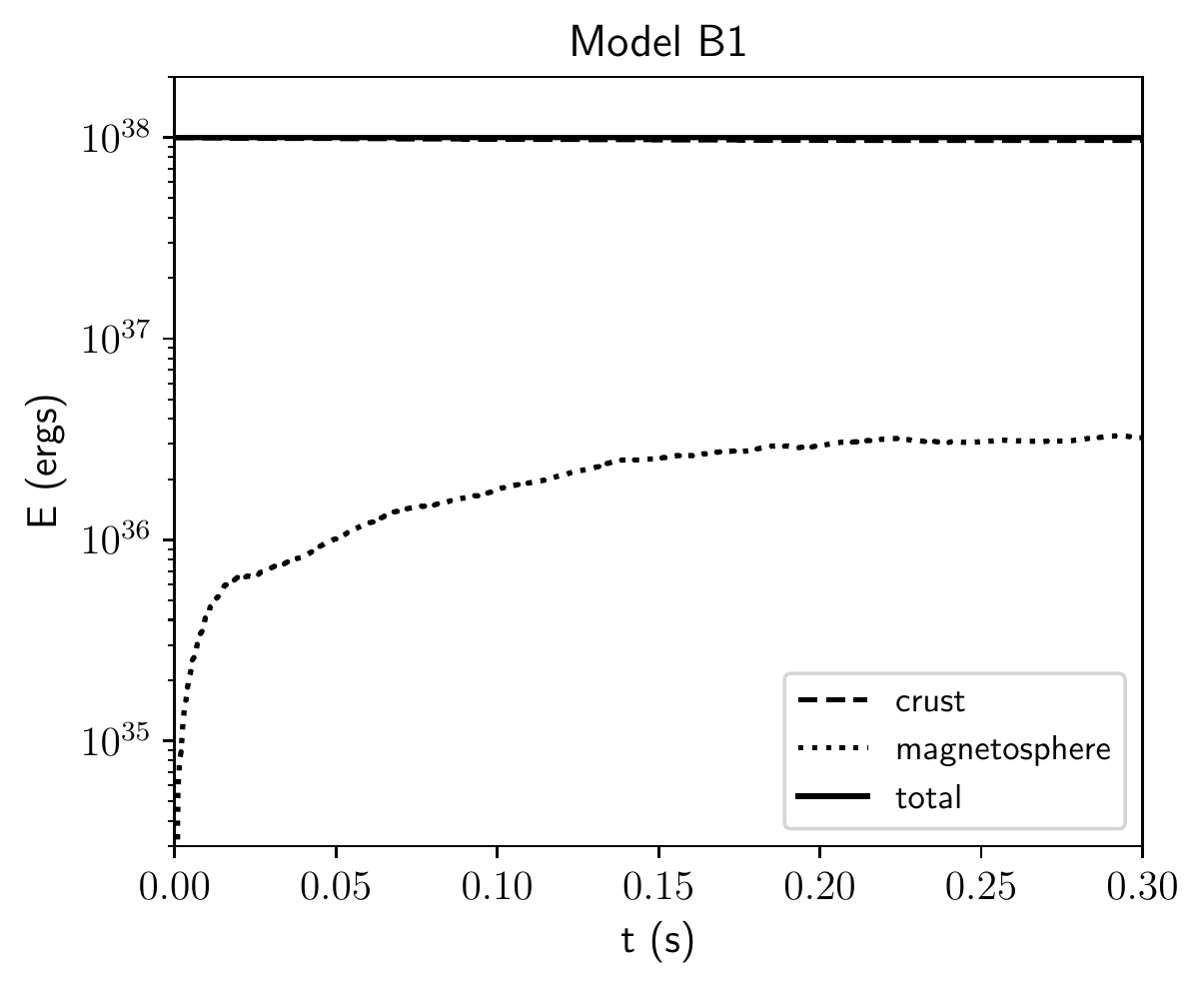}
\includegraphics[width=.40\textwidth]{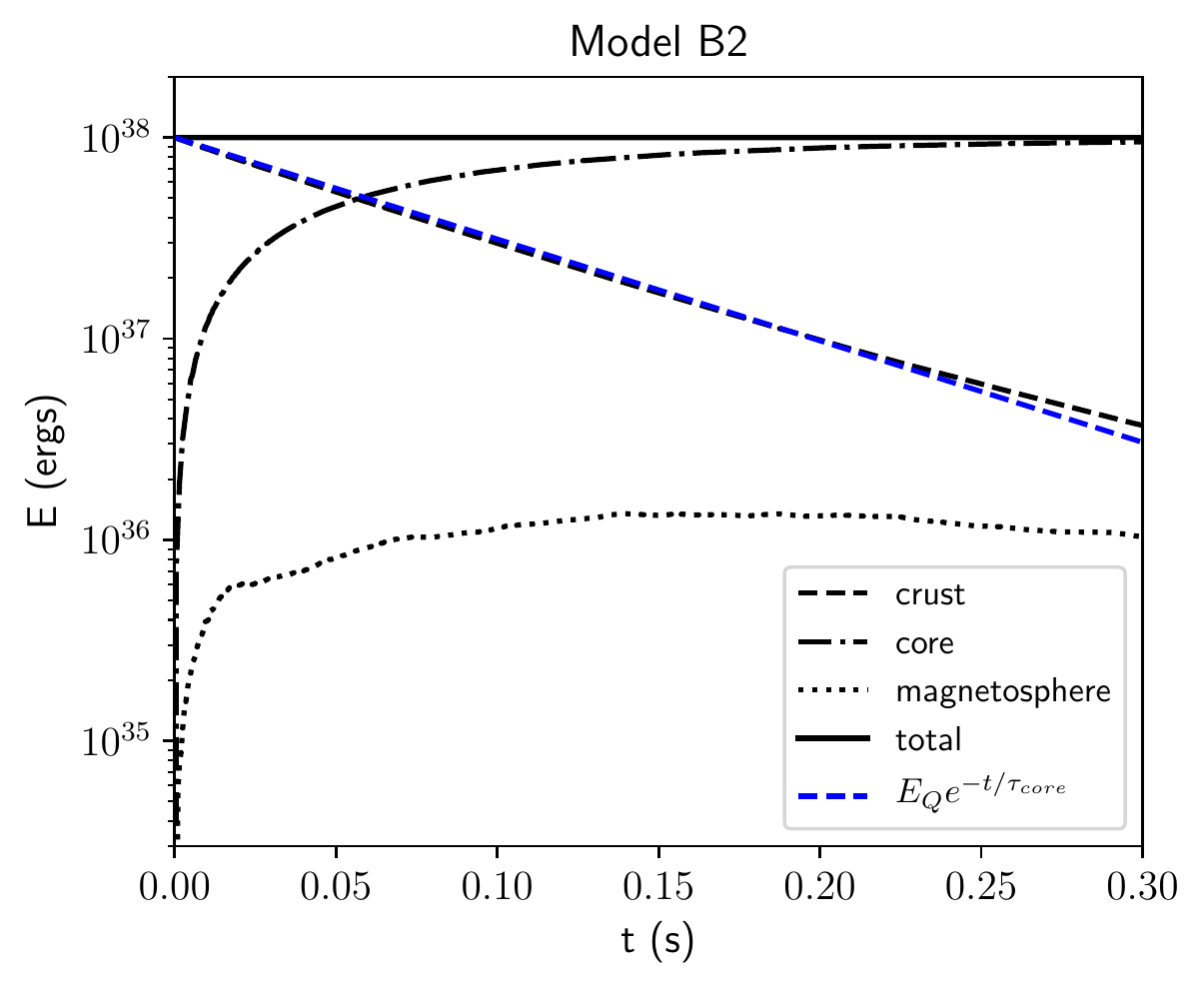}
\caption{
Evolution of the quake energy. The four panels show the results for models A1, A2, B1 and B2. The energy retained by the crustal oscillations $E_{\rm crust}$ (dashed curve) is reduced by the transmission into the magnetosphere (dotted) and (in models A2, B2) transmission into the core (dotted-dashed). As required by energy conservation, the sum of the retained and transmitted energies remains equal to $E_Q=10^{38}\,$erg (horizontal solid line). The blue dashed line shows the analytical approximation to $E_{\rm crust}(t)$ (Equation~\ref{eq:Ecrust}) with $\tau_\text{core}=86\,$ms.
}
\label{energy}
\end{figure*}

The magnetospheric Alfv\'en waves are initially coherent when launched from the surface (Figure \ref{A1}, top right), with the perpendicular length scale determined by the length of the elastic waves in the crust. After a light-crossing time ($\sim 45$ ms for the last closed field-line) all of the Alfv\'en waves become de-phased (Figure \ref{A2}, top right). The regions where $|j_\parallel/c\rho_{\rm GJ}|>1$ are mapped in Figures~\ref{A1} and \ref{A2}. We find that avoiding charge starvation and the ignition of $e^\pm$ discharge requires the magnetospheric plasma to have a high multiplicity $\M\gtrsim 10^3$, in agreement with the estimates in Section~\ref{discharge}. After three rotations of Vela, $L_{\rm A}$ has dropped by a factor of $\sim2-3$. Less than $3$\% of the quake energy $E_Q$ has been transferred to the magnetosphere (Figure \ref{energy}). 

\begin{figure*}
\centering
\includegraphics[height=0.309\textwidth]{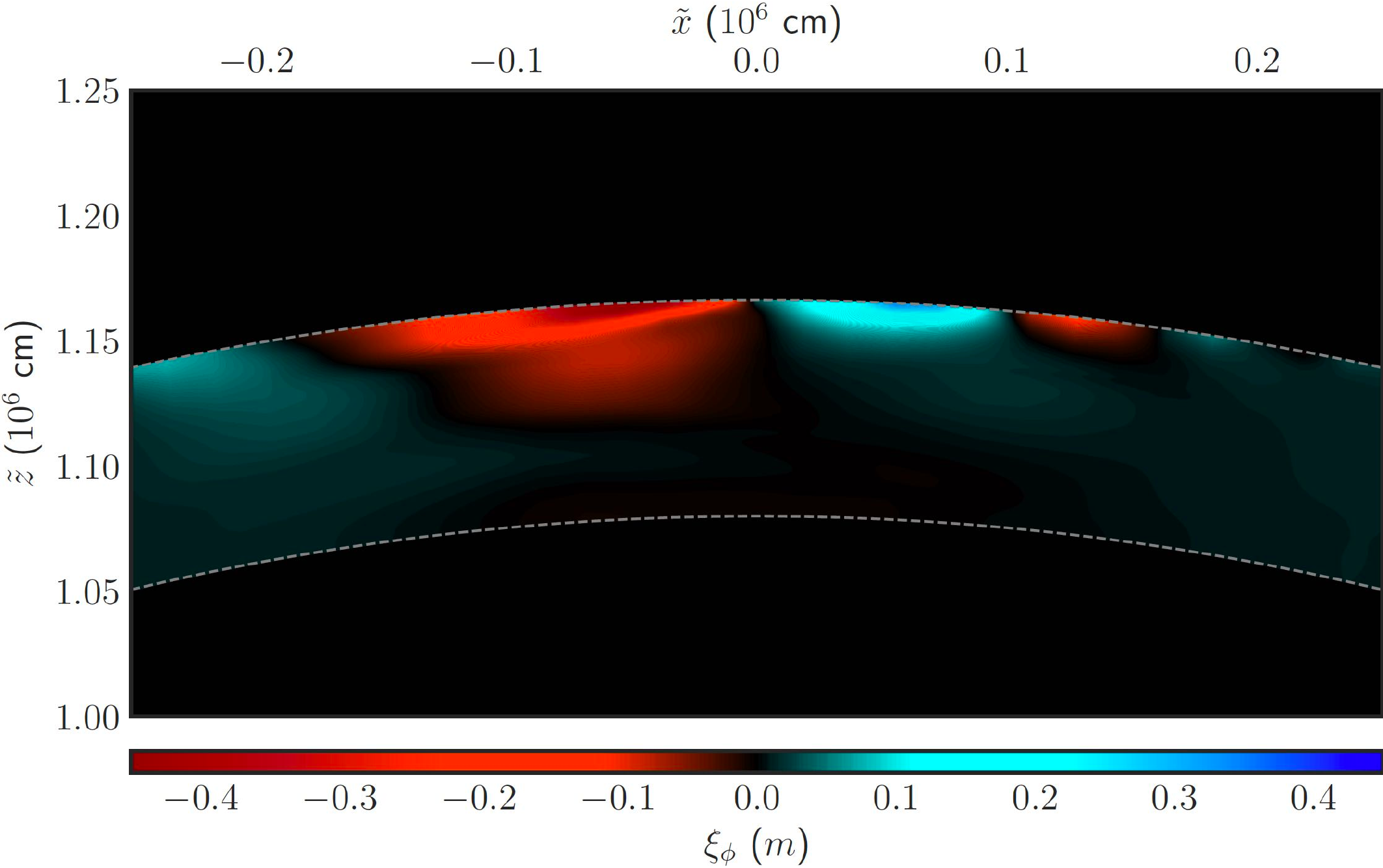} 
\includegraphics[height=0.31\textwidth]{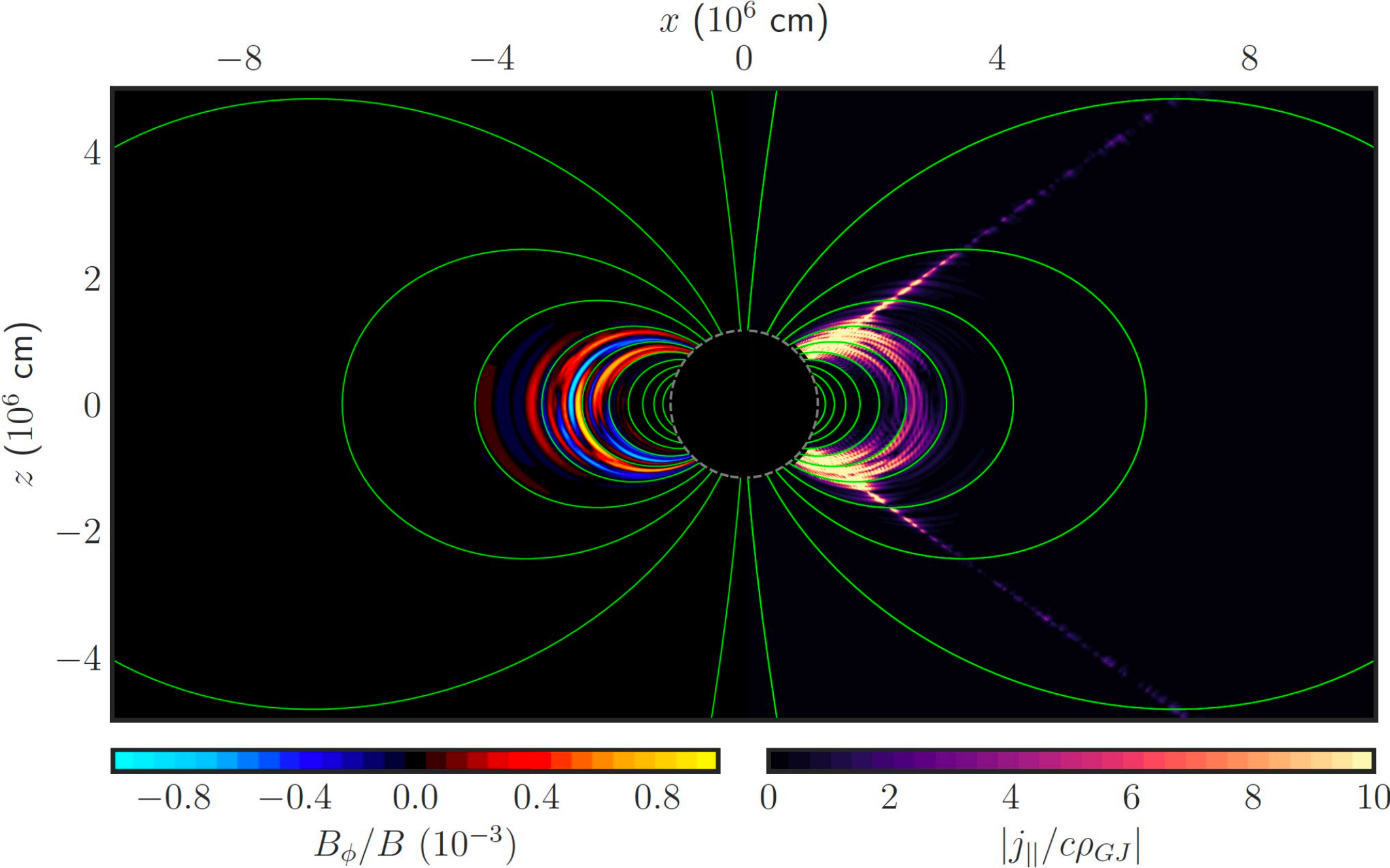} 
\hspace*{-.11in}
\includegraphics[width=1.01\textwidth]{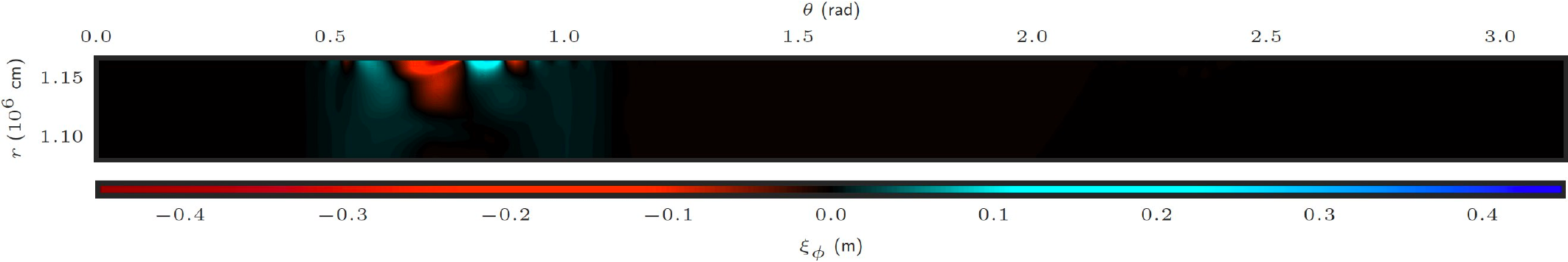} 
    \caption{Model B1 at $t=2$ ms. Top left: displacement $\xi_\phi$ of the crust near the epicenter of the quake. The epicenter is at $\theta=45$~$^{\circ}$, and we have rotated the figure by $-45$~$^{\circ}$ ($\tilde{x}=x-z$ and $\tilde{z}=x+z$). The dashed lines show the boundaries of the crust. Top right: toroidal perturbation of the magnetic field $B_\phi / B$ (left), and the ratio $|j_\parallel / c \rho_{GJ}|$ (right). The green curves show the poloidal magnetic field. The two field lines closest to the axis of symmetry are the boundary of the open field-line bundle. The gray dashed circle is the surface of the neutron star. Bottom: displacement $\xi_\phi(r,\theta)$ in the entire crust, plotted on the $r$-$\theta$ plane.}
    \label{C1}
\end{figure*}

\begin{figure*}
\centering
\includegraphics[height=0.309\textwidth]{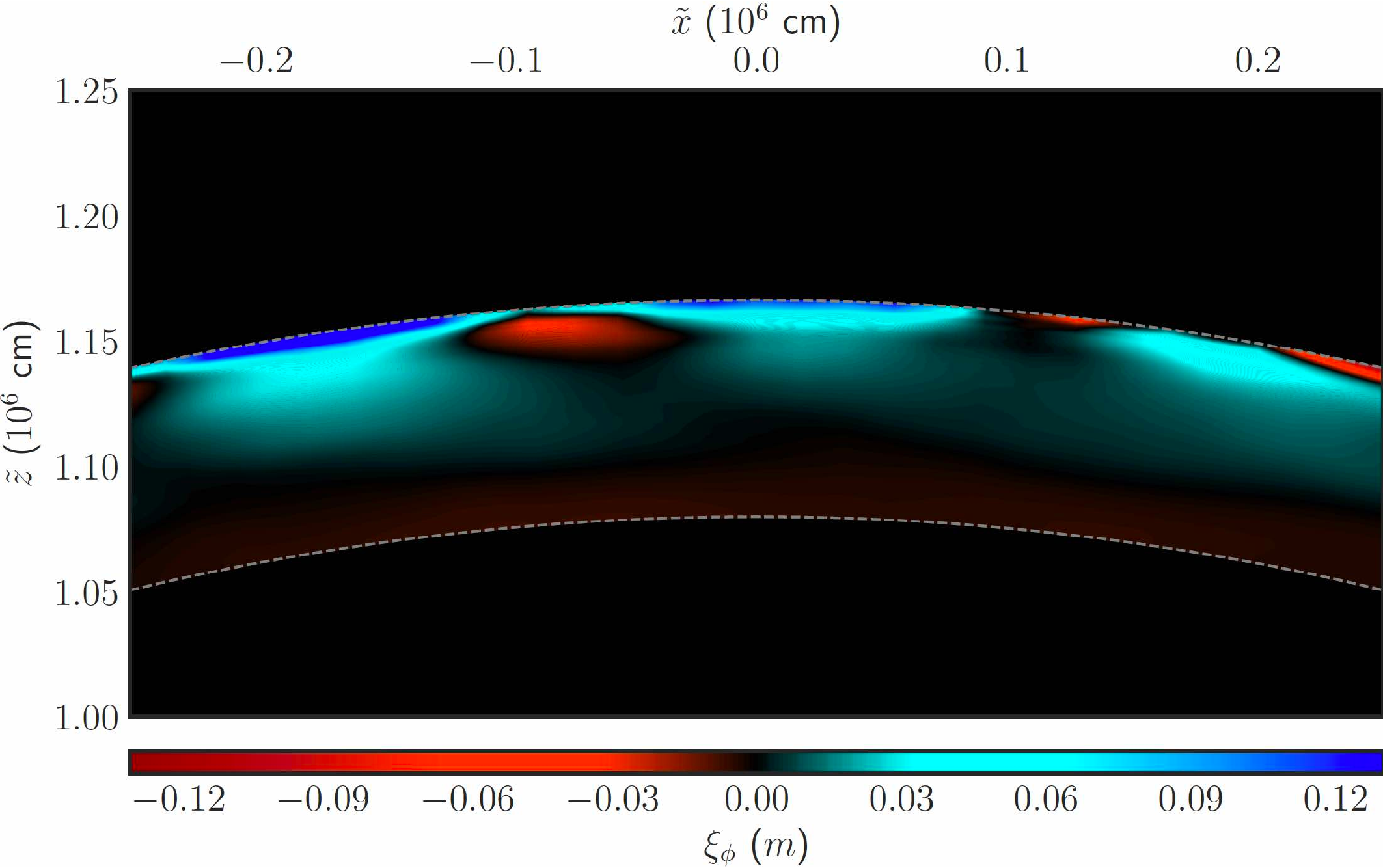} 
\includegraphics[height=0.31\textwidth]{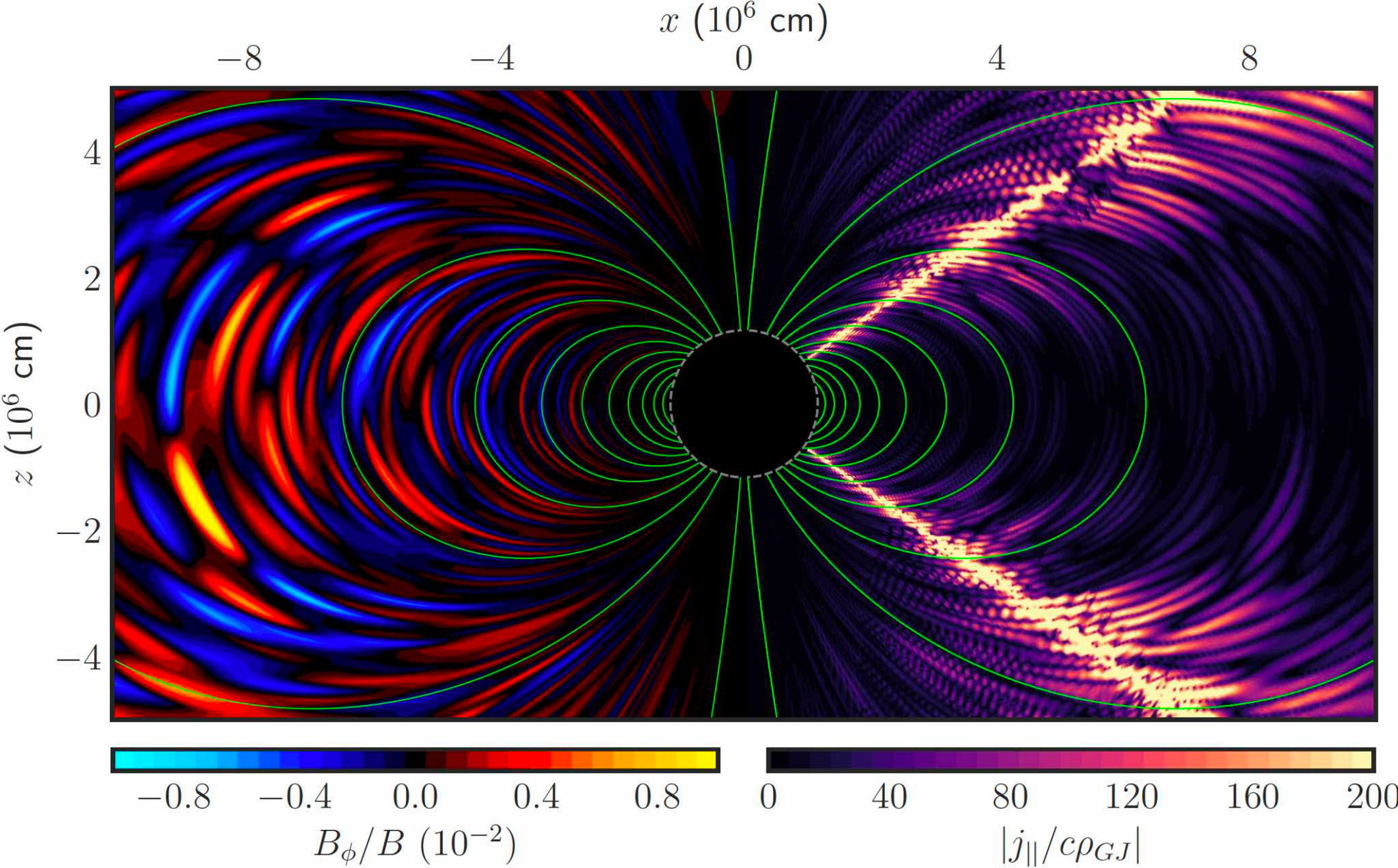} 
\hspace*{-.18in}
\includegraphics[width=1.01\textwidth]{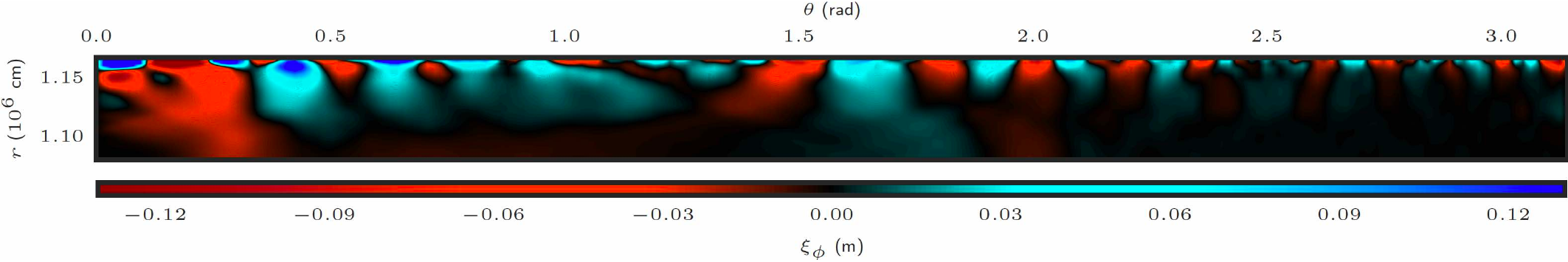} 
    \caption{Same as Figure~\ref{C1} but at time $t=50\,$ms.}
    \label{C2}
\end{figure*}

The dynamics in model~B1 is the same except that the elastic waves spread from a different quake region, now located at latitude $\theta\sim \pi/4$ instead of the north pole (Figures \ref{C1} and \ref{C2}). The energy budget and the timescale for injecting the Alfv\'en waves into the magnetosphere are similar to those in model~A1.
At first, Alfv\'en waves are only launched into the closed field-lines (Figure \ref{C1}), but after $\sim 20$ ms the crustal shear waves have spread to the north polar cap, and Alfv\'en waves are launched into the north open field-line bundle, and the entire closed magnetosphere. Their luminosity $L_{\rm A}$ remains quite constant for the remainder of the simulation. After 3 rotations of Vela, $\sim 3$\% of the initial elastic energy has been transmitted into the magnetosphere.

Models~A2 and B2, which include the crust-core coupling, show a significant difference from models A1 and B1: the lifetime of crustal waves is significantly reduced, because the wave energy is drained into the core. This draining occurs exponentially, because it results from the damping force $f_{\rm core}\propto \dot{\xi}$ (Equation~\ref{f_core}). The evolution of the crustal wave energy is well approximated by
\beq
\label{eq:Ecrust}
   E_{\rm crust}\approx E_Q \exp\left(-\frac{t}{\tau_{\rm core}}\right),
\eeq
with $\tau_{\rm core}\approx 86\,$ms in both models A2 and B2 (Figure~\ref{energy}). The luminosity of Alfv\'en waves into the magnetosphere $L_{\rm A}$ decays on the same characteristic timescale. After three rotations of Vela, $\sim 1$\% of the initial elastic energy is in the magnetosphere, and $\sim 95$\% of the initial energy has been transmitted into the liquid core. The luminosity $L_{\rm A}$ has decreased by a factor of $\sim 20$. The evolution of $L_{\rm A}$ and the wave energy in all four models is summarized in Figures~\ref{lum} and \ref{energy}.



\section{Discussion}\label{discussion}
Glitches give deep insight into the exotic dynamics of quantum fluids that likely exist in pulsar interiors. One of the unsolved theoretical issues is the cause of the nearly simultaneous unpinning of billions of superfluid vortices over a macroscopic $10-10^3$ m length that must take place during a glitch. The catastrophic unpinning is required to explain the glitches' magnitudes, especially the giant glitches with the relative spin-up of $\sim 10^{-5}$ observed in Vela. Crustal quakes have been suggested as one of the candidates for the glitch trigger, but not considered promising for Vela. Indeed, what could deform the crust so dramatically that it would have a mechanical failure? Vela's external magnetic field is two orders of magnitude smaller than that of magnetars, and thus the magnetic stresses are not obviously sufficient to break the crust. Furthermore, Vela is spinning at 1\% of the break-up  angular velocity, and thus its relative rotational deformation is $\sim 10^{-4}$, which is smaller than the critical strain of the crust. Therefore, rotational deformation is also unlikely to lead to a quake.

Nonetheless, the remarkable observations of the 2016 glitch by \cite{palfreyman_alteration_2018} force one to seriously consider a quake as a trigger. The change in the magnetospheric activity indicates its strong disturbance by the glitch on a timescale shorter than 0.1~s. The only plausible way for such a disturbance to be delivered from the star's interior is through a shear wave that reaches the interface between the crust and the magnetosphere. The high-frequency elastic wave can shake vortex pinning sites in the crust. The resulting magnus force can unpin vortices in a macroscopic region causing a glitch \citep{eichler_dynamical_2010}. Alternatively, plastic failure of the crust could generate sufficient heat to thermally unpin many vortices \citep{link_thermally_1996}.

In this paper, we studied an important ingredient of such a scenario --- the seismic motion in the crust and its coupling to the magnetosphere and the core. We have shown that the seismic activity, once created, spreads through the crust and engages the whole magnetosphere in Alfv\'en-type oscillations. Even for a modest-amplitude quake, we find that the magnetospheric disturbance can cause an electric discharge that produces gamma-rays and $e^\pm$ pairs. We are unable to make specific predictions for the quake effect on the radio luminosity $L_{\rm GHz}$, because the mechanism of pulsar emission is poorly understood. However, it is reasonable to expect that the appearance of a new powerful $e^\pm$ source changes $L_{\rm GHz}$ for the duration of the quake, and could shut down the radio pulsations as observed in the Vela glitch in 2016 December. The seismic motion in the crust is damped on a short timescale through  emission of Alfv\'en waves into the liquid core. This process is sped up by the enhanced magnetic tension due to the bunching of the magnetic field into flux tubes in the superconducting core of Vela. As a result, the damping timescale for the crustal oscillations is as short as $\sim 0.2$ s, comparable to the duration of the observed pulse disturbance.

New detailed observations would help confirm the presence of magnetospheric disturbances during glitches. If such disturbances turn out to be common, they will require a paradigm shift that should include crustal quakes as a common phenomenon in young pulsars. This could indicate internal magnetic fields that are orders of magnitude greater than the external dipole component responsible for the pulsar spin-down. The existence of ultra-strong internal fields would not require the assumption of superconductivity to explain the short lifetime of the quake. In addition, it would indicate that the Vela glitches are due to the crustal superfluid, contrary to models that invoke the core superfluid [e.g. \cite{ruderman_neutron_1998},  \cite{sidery_effect_2009}]. The theoretical challenges pertaining to pulsar exteriors would also be considerable: the damping of the strong magnetospheric waves and their impact on pair production and pulsar radio emission will need to be understood.

The methodology developed in this paper is not limited to studies of quakes in pulsars, but can also be used for studies of magnetars, where superstrong crustal quakes were proposed as triggers of giant X-ray flares \citep{thompson_soft_1996}.  

Finally, we note that the quake we invoked for the Vela glitch is capable of producing a weak X-ray burst. We found the Alfv\'en wave energy deposited in the magnetosphere $E_{\rm A}\sim 10^{-2}E_Q\sim 10^{36}\,$erg. This energy is dissipated through the discharge, and a large fraction of $E_{\rm A}$ should be emitted in the X-ray band. In particular, X-rays are emitted by $e^\pm$ created near the star in excited Landau states, and cascading down to the ground state. The duration of the X-ray burst is comparable to the dissipation timescale for the magnetospheric Alfv\'en waves. The burst is much brighter than than the normal pulsating X-ray luminosity of Vela; however, its detection is challenging because of the short duration and the modest fluence.

\section{Acknowledgments}
We thank Mal Ruderman, Andrei Gruzinov, and Xinyu Li for useful discussions. A.M.B. is supported by NASA grant
NNX17AK37G, a Simons Investigator Award (grant
No. 446228), and the Humboldt Foundation.

\appendix

\section{Elastic modes}\label{elastic_modes}
The elastic modes $\ff_{nl}(r)$ and corresponding frequencies $\omega_{nl}$ are found by solving the eigenvalue equation [Equation \eqref{f_eqn}],
\begin{equation}
-\omega_{nl}^2 \rho \ff_{nl} =  \frac{d\tilde{\mu} }{dr}\bigg(\frac{d\ff_{nl}}{dr} - \frac{\ff_{nl}}{r} \bigg) + \frac{\tilde{\mu}}{r^2}\frac{d}{dr}\bigg(r^2\frac{d\ff_{nl}}{dr}\bigg) -[ l(l+1)\mu + 2\mu_B]\frac{\ff_{nl}}{r^2}.
\end{equation}
Following \cite{mcdermott_nonradial_1988} Equation \eqref{f_eqn} is reduced to two first-order ordinary diffrential equations by introducing the dimensionless variables 
\begin{equation}
S_1\equiv \frac{\ff_{nl}}{r},
\end{equation}
\begin{equation}
S_2 \equiv \frac{\tilde{\mu} r_\star}{\omega^2M_\star}\left( \frac{d \ff_{nl}}{dr}-\frac{\ff_{nl}}{r}\right),
\end{equation}
where $S_1$ refers to a dimensionless amplitude, and $S_2$ is a dimensionless stress. In terms of these variables, the equation for $\ff_{nl}$ becomes
\begin{equation}
r\frac{d S_1}{dr} = \frac{\omega^2}{\tilde{\mu}}\frac{M_\star}{r_\star}S_2,
\label{S1}
\end{equation}
\begin{equation}
r\frac{dS_2}{dr} = \frac{\mu r_\star}{\omega^2 M_\star}\left[ l(l+1) -2 -\frac{\omega^2 \rho r^2}{\mu} \right]S_1 -3S_2. 
\label{S2}
\end{equation}
In the limit $\mu_B\longrightarrow0$ Equations~\eqref{S1} and \eqref{S2} reduce to Equations~25(a) and (b) of \cite{mcdermott_nonradial_1988}. The appropriate boundary conditions for these unforced modes are zero magnetic stress $\sigma_{r\theta}^{\text{mag}}=\sigma_{r\phi}^{\text{mag}}=0$ and zero elastic stress $\sigma_{r\theta}^{\text{el}}=\sigma_{r\phi}^{\text{el}}=0$ at the boundaries. These conditions are expressed through the single equation
\begin{equation}
\tilde{\mu}\left( \frac{d\ff_{nl}}{dr}-\frac{\ff_{nl}}{r}\right) = 0,
\end{equation}
or in terms of the variable $S_2$,
\begin{equation}
S_2(r_i)=0,
\label{bc1}
\end{equation}
where $r_i$ is either the radius of the crust-core interface ($r_c$), or the surface of the crust ($r_\star$). The amplitude of the displacement is arbitrary, as the problem is linear. We set the amplitude at the crust-core interface
\begin{equation}
S_1(r_c)=1.
\label{bc2}
\end{equation}
Equations \eqref{S1} and \eqref{S2}, together with the boundary conditions Equations \eqref{bc1} and \eqref{bc2}, constitute a well posed Sturm-Liouville problem. 

The Sturm-Lioville problem is solved by `shooting' (integrating) from the crust-core interface and varying the eigenvalue until the boundary condition Equation \eqref{bc1} is satisfied at the surface of the crust. We have implemented a fourth-order Runge-Kutta integrator, and used it in two modes: i) Scanning: for each value of $l$ the eigenvalue is varied coarsely through all possible values up to some maximum frequency. The frequencies for which $S_2(r_\star)$ is minimized are recorded as estimates of the eigenvalues, together with the corresponding value of $n$. ii) Root finding: for each $(n,l)$ Newton-Raphson method is used to converge on the eigenvalue $\omega_{nl}$ for which $|S_2(r_\star)|< \epsilon_\star$ (typically we set $\epsilon_\star=10^{-12}$). The frequencies from the scanning mode are used as first guesses for the Newton-Raphson iterations.

When finding modes we use a uniform radial grid of $50,000$ points. As a test we check the orthogonality of our modes. We typically find
\begin{equation}
\int_{r_c}^{r_\star} \rho r^2 \ff_{nl}\ff_{n'l}dr = \delta_{nn'}\pm 10^{-9}. 
\end{equation}
We also studied the time-dependent propagation of a radial $l=0$ wave using our elastic modes. This was compared to the same wave propagation using a 1D finite difference solver. The two methods produced the same time-dependent solution. To test the convergence, we found one set of modes on a grid of $20,000$ points, and another on a grid of $50,000$ points. We ran simulations of 2D axisymmetric elastic waves with both sets of modes, using the same initial conditions. The time-dependent solutions were indistinguishable, indicating that our elastic modes and frequencies are converged to a sufficient accuracy for our dynamical simulations. The obtained normalized modes $\ff_{nl}$ and their frequencies $\omega_{nl}$ are stored and used for the dynamical simulations described below.

\section{Crust Dynamics: Numerical method}\label{numerical_method_crust}
The spectral method follows the dynamics of the crust through the coefficients $a_{nlm}(t)$. Since we are only considering axisymmetric dynamics in this work, the index $m$ is set to zero, and $\xi_\phi$ is the only nonzero component of the displacement. The displacement is written as a sum over basis functions (orthogonal eigenmodes),
\begin{equation}
\vec{\xi}(t,r,\theta) = \xi_\phi(t,r,\theta)\vec{\hat{\phi}} = \sum_{n=0}^{n_\text{max}} \sum_{l=1}^{l_\text{max}} a_{nl}(t)\vec{\xi}_{nl}(r,\theta),
\label{displacement}
\end{equation}
where finite $n_\text{max}$ and $l_\text{max}$ are chosen to truncate the infinite series. The product $n_\text{max}\times l_\text{max}$ is the total number of the eigenmodes in our simulations. The basis functions are 
\begin{equation}
\vec{\xi}_{nl} = \xi^\phi_{nl}\vec{\hat{\phi}} = \ff_{nl}(r)\frac{d Y_{l0}(\theta)}{d\theta}\vec{\hat{\phi}},
\end{equation}
where $Y_{l0}=P_l(\cos\theta)$ are the Legendre polynomials and the radial eigenfunctions $\ff_{nl}(r)$ are found as described in Appendix~\ref{elastic_modes}. The initial conditions are set by projecting $\vec{\xi}(t=0)$ on to the basis functions $\vec{\xi}_{nl}$ for each $(n,l)$,
\begin{equation}
a_{nl}(t=0) = \langle\vec{\xi}(\vec{r},t=0),\vec{\xi}_{nl}\rangle=\int_{r_c}^{r_\star}dr\int_{0}^{\pi}d\theta \, r^2 \sin\theta\,\rho\, \xi_\phi(t=0)\, \xi^\phi_{nl},
\end{equation}
where we have used that the modes are orthonormal. The integration is done numerically on a uniform $(r,\theta)$ grid of $N_r\times N_\theta =1000\times600$ points using the fifth-order accurate Simpsons rule. The Legendre polynomials $P_l$ and derivatives are computed once at the beginning of the simulation and stored. The time evolution of $a_{nl}$ is given by the equation of motion
\begin{equation}
\ddot{a}_{nl}(t) + \omega_{nl}^2 a_{nl}(t) = \langle \vec{f}_{\text{ext}}(\vec{r},t),\vec{\xi}_{nl}\rangle,
\label{equation_of_motion}
\end{equation}
where $ \vec{f}_{\text{ext}}= \vec{f}_{\text{core}}+ \vec{f}_{\text{mag}}$ is the force on the crust due to the core and magnetosphere, and $\langle \vec{f}_{\text{ext}}(\vec{r},t),\vec{\xi}_{nl}\rangle$ is a matrix containing the projection of  $ \vec{f}_{\text{ext}} $ onto the basis functions. The force of the core on the crust (Equation \eqref{f_core}) is written as
\begin{equation}
 \vec{f}_{\text{core}} = -v_A\frac{\rho_<}{\rho_>}\delta(r-r_c)\,\dot{\vec{\xi}} =  -v_A\frac{\rho_<}{\rho_>}\delta(r-r_c) \sum_{n=0}^{n_\text{max}} \sum_{l=1}^{l_\text{max}} \dot{a}_{nl}(t)\vec{\xi}_{nl} ,
\end{equation}
where we have used Equation \eqref{displacement} to express the $\dot{\vec{\xi}}$ in terms of the coefficients $\dot{a}_{nl}$. Then the projection of $\vec{f}_{\text{core}}$ onto the basis functions is given by
\begin{equation}
\langle \vec{f}_{\text{core}}(\vec{r},t),\vec{\xi}_{nl}\rangle =  \sum_{n'=0}^{n_\text{max}} \sum_{l'=1}^{l_\text{max}} \dot{a}_{n'l'}(t)C_{n'l'nl},
\label{core_force}
\end{equation}
where 
\begin{equation}
C_{n'l'nl}= -\int_{r_c}^{r_\star}dr\int_{0}^{\pi}d\theta r^2 \sin\theta \rho v_A\frac{\rho_<}{\rho_>} \delta(r-r_c)\xi^{\phi}_{n'l'}\xi^{\phi}_{nl} =  -r_c^2 v_A \rho_<  f_{n'l'}(r_c) f_{nl}(r_c)\delta_{ll'}.
\end{equation}
The components of the coupling matrix $C_{n'l'nl}=0$ for $l'\neq l$; therefore,
it is not necessary to sum over $l'$ in Equation \eqref{core_force}. The matrix $C_{n'l'nl}$ is calculated once at the beginning of each simulation and stored.

The force of the magnetosphere on the crust is 
\begin{equation}
f^{\phi}_\text{mag} = \frac{\rho_B}{\rho_{\rm crys}}c^2 \cos\alpha \, \delta(r-r_\star)\,r_\perp\frac{\partial}{\partial\chi}\left( \frac{\xi_\phi}{r_\perp}\right)\bigg|_{r_>}.
\label{force_mag}
\end{equation}
As $\xi_\phi$ is evolved self-consistently in the magnetosphere (Appendix \ref{numerical_method_magnetosphere}), the force $f^{\phi}_\text{mag}$ is calculated at each time step and used to evaluate
\begin{equation}
\langle \vec{f}_{\text{mag}}(\vec{r},t),\vec{\xi}_{nl}\rangle =   \int_{r_c}^{r_\star}dr\int_{0}^{\pi}d\theta r^2 \sin\theta\, \rho\, f^{\phi}_\text{mag}(t,r,\theta)\, \xi^{\phi}_{nl} = r_\star^2 \rho(r_\star)\ff_{nl}(r_\star) \int_{0}^{\pi}d\theta \sin\theta f^{\phi}_\text{mag}(t,r_\star,\theta) \frac{\partial Y_{l0}}{\partial \theta} ,
\label{mag_force}
\end{equation}
where the integral is evaluated numerically at each time step on a uniform grid of $N_\theta$ points using the fifth-order Simpsons rule. Equation \eqref{equation_of_motion} is integrated in time together with using the fourth-order Runge-Kutta integration, with a constant time step $\Delta t=\text{min}\lbrace\Delta t_\text{crust},\Delta t_\text{mag}\rbrace$, where $\Delta t_\text{crust}$ is the largest stable time step for the crust, and $\Delta t_\text{core}$ is the largest stable time step for the magnetosphere (see Appendix \ref{numerical_method_magnetosphere}). We use $\Delta t_\text{crust} = k_c/\text{max}\lbrace\omega_{nl}\rbrace$ with $k_c\leq 0.1$, where $\text{max}\lbrace\omega_{nl}\rbrace$ is the highest frequency of all of the modes we are using. We have found that for a free crust (without external forcing terms), our code conserves energy to one part per million. If the external forcing terms are included, some additional error is introduced, and energy is usually conserved to one part in $10^5$. 

We use $(n_\text{max},l_\text{max})=(300,200)$, a total of 60,000 modes. More radial modes are needed ($n_\text{max}>l_\text{max}$) to properly resolve the wave transmission through the upper layers of the crust where the scale height is very small. The only relevant scale in the $\theta$-direction is introduced by the initial conditions. We have tried independently increasing $n_\text{max}$ to 600, and $l_\text{max}$ to 400, and we observe the same results. 

\section{Magnetosphere dynamics: Numerical method}\label{numerical_method_magnetosphere}
In the magnetosphere, we calculate the small azimuthal displacement $\xi_\phi$, using the so-called magnetic flux coordinates $(\psi,\chi,\phi)$, where 
$\psi=$const
defines surfaces of constant poloidal flux, and $\chi$ is the length along poloidal 
the
field lines in the $\phi=\text{const}$ plane. The dependence of the Cartesian position vector $\boldsymbol{x}$ on the coordinates $\psi$ and $\chi$ is found by integrating the equation
\begin{equation}
\frac{d\vec{x}(\psi,\chi)}{d\chi} = \frac{\vec{B}}{|\vec{B}|}.
\end{equation}
The footpoints of the field lines are chosen to coincide with the grid points used in the projection Equation \eqref{mag_force}. We chose the grid spacing along 
the
field lines so that the light-crossing time of each grid cell is the same. When we include the liquid ocean, the grid spacing remains large in the magnetosphere, but becomes very small in the 
ocean where the density increases. By using this grid spacing, we are not limited to a prohibitively small time step by the Courant condition. The time evolution of $\xi_\phi (\psi,\chi)$ is given by the wave equation
\begin{equation}
\frac{\partial^2 \xi_\phi(\psi,\chi)}{\partial t^2} = \frac{B}{4\pi r_\perp \rho_B} \frac{\partial}{\partial \chi}\left[ r_\perp^2 B \frac{\partial}{\partial \chi}\left(\frac{\xi_\phi(\psi,\chi)}{r_\perp}\right)\right]. 
\label{alfven_eqn}
\end{equation}
We are effectively solving a 1D wave equation for each flux surface $\psi$. The right-hand side of Equation \eqref{alfven_eqn} is evaluated using the second-order finite difference formulas given by \cite{bowen_derivative_2005}. The first derivatives use a three-point stencil, and the second derivatives use a four-point stencil, so that second-order accuracy is preserved when the grid spacing is nonuniform. We integrate Equation \eqref{alfven_eqn} in time, together with Equation \eqref{equation_of_motion} for the crust using the fourth-order Runge-Kutta integration. 

The crust provides the boundary condition for $\xi_\phi(\psi,\chi)$ at the surface in the magnetosphere,  and the magnetosphere communicates to the crust through the force Equation \eqref{force_mag}. The stable time step for the magnetosphere is $\Delta t_\text{mag}=k_c dt_\chi$, where $dt_\chi$ is the light-crossing time of a grid cell, and $k_c<0.5$. We set the time step for the simulation $\Delta t=\text{min}\lbrace\Delta t_\text{crust},\Delta t_\text{mag}\rbrace$, where $\Delta t_\text{crust}$ is the largest stable time step for the crust (see Appendix \ref{numerical_method_crust}). We find that $\sim 600$ grid points are required for the projection Equation \eqref{mag_force}, which results in $\sim 50$ open flux surfaces ($\sim 25$ at each pole), and $\sim 275$ closed flux surfaces.

\section{A Test for Wave Transmission}\label{numerical_transmission}
\begin{figure*}[h!]
\centering
\includegraphics[width=.46\textwidth]{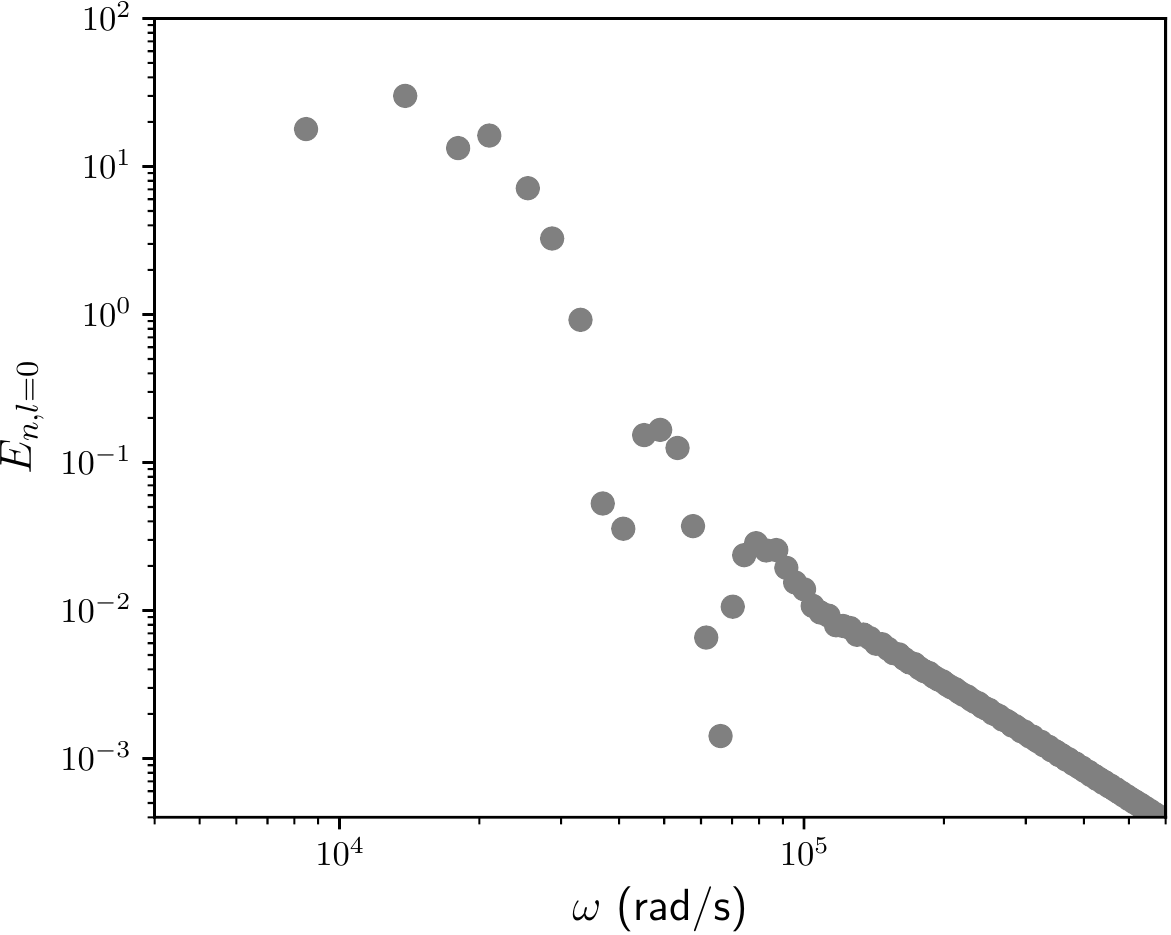}
\includegraphics[width=.46\textwidth]{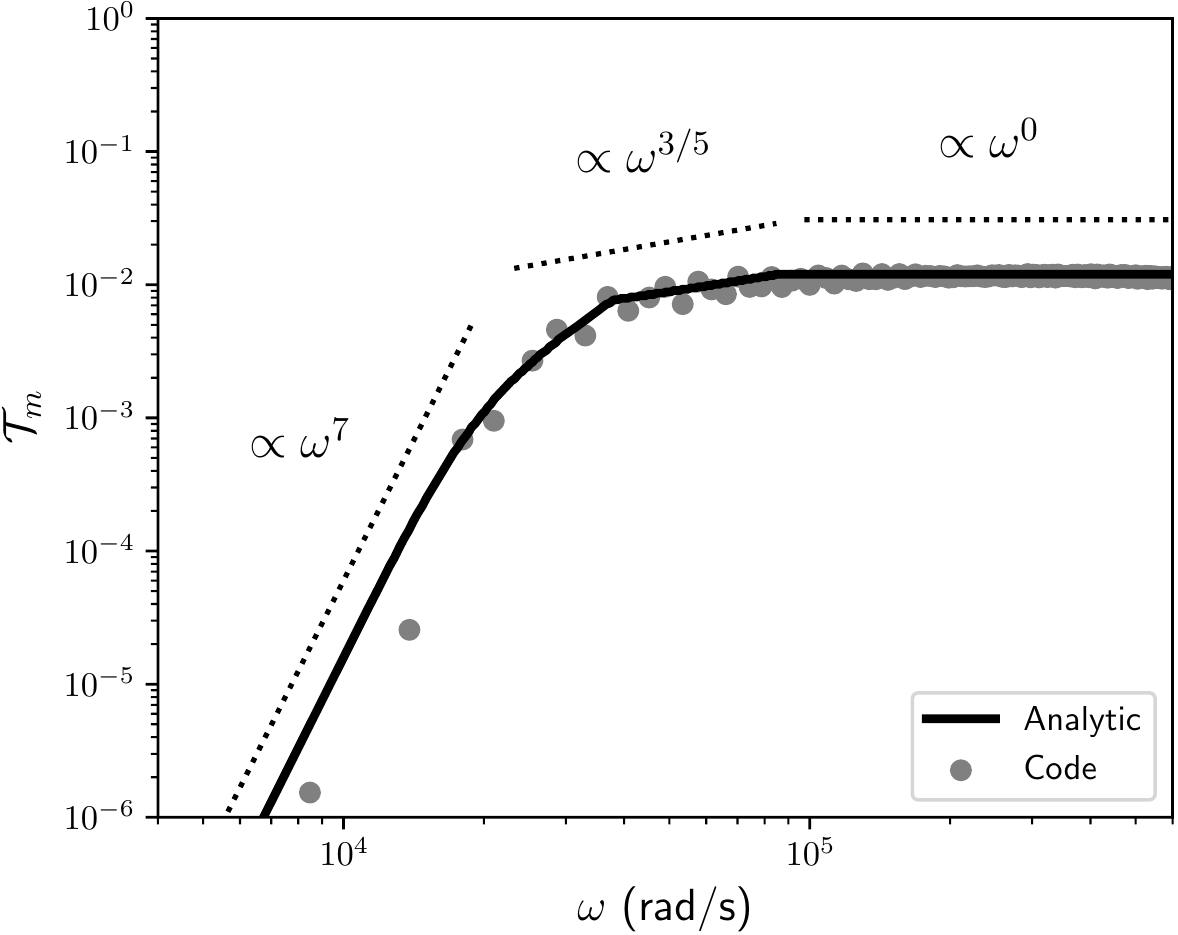}
\caption{Left: initial energy spectrum of waves in this test problem. Right: transmission coefficient of waves into the magnetosphere $\T_m(\omega)$. The thick black line shows the transmission coefficient found by solving the analytic reflection conditions (Section \ref{quake}), and the gray dots show the numerical transmission coefficient measured using our code for an $l=0$ radial wave. The dotted lines show the powe-law scalings in each frequency range.}
\label{trans}
\end{figure*}
In order to test the implementation of the crust-magnetosphere coupling in our numerical model, we have measured the frequency-dependent transmission coefficient $\T_m(\omega)$ using our code. We initialize the simulation by launching a purely radial $l=0$ wave in the crust. The magnetosphere is chosen to be a radial monopole with outflow boundary conditions on \emph{all} flux surfaces, so that no Alfv\'en waves return to the crust. The setup is effectively 1D, so that we should recover the transmission coefficient calculated in Section \ref{quake} for a Cartesian slab crust. 

The energy spectrum of the initial condition is shown in Figure \ref{trans} (left panel). The initial displacement is a smoothed step function, similar to the 2D initial conditions used in Section \ref{results}. It corresponds to a strain layer of thickness $\Delta\ell \sim 10^4$ cm, similar to the pressure scale height in the deep crust. The energy spectrum peaks around $\omega \sim \tilde{v}_s/\lQ \sim 2\times 10^4$ rad/s. 

We measure the transmission coefficient by calculating the exponential decay time of the energy in each mode $\tau_m$. The transmission coefficient for a given elastic mode is then calculated as $\T_m = 2\tau / \tau_m$, where $\tau\approx 1$ ms is the elastic wave crossing time of the crust. This gives the effective transmission coefficient as a function of the mode frequency, $\T_m(\omega)$, which we compare with the analytically calculated $\T_m$ (Figure \ref{trans}). The two lowest-frequency modes deviate from the analytical result, because they are reflected deep in the crust near neutron drip, where the exact density profile used in the code deviates from the approximation $\rho\propto |z|^3$ used in the analytical model. There are few data points at low frequencies in Figure \ref{trans} because there are few elastic modes in that frequency range. We performed similar simulations with different initial conditions and found nearly the same $\T_m(\omega)$.

\bibliographystyle{apj}
\bibliography{Astrophysics}



\end{document}